\documentclass[12pt,preprint]{aastex}
\shorttitle{On the migration of protogiant solid cores}
\shortauthors{F.\ S.\ Masset, G.\ D'Angelo \& W.\ Kley}
\begin{document}
\title{On  the migration  of protogiant  solid cores}  \author{F.\ S.\
  Masset\altaffilmark{1,2}}      \affil{AIM      -      UMR      7158,
  CEA/CNRS/Univ. Paris  7, SAp,  Orme des Merisiers,  CE-Saclay, 91191
  Gif/Yvette Cedex,  France} \altaffiltext{1}{Also at  IA-UNAM, Ciudad
  Universitaria,  Apartado  Postal 70-264,  Mexico  DF 04510,  Mexico}
\email{fmasset@cea.fr}
\author{G. D'Angelo\altaffilmark{3}}\affil{NASA-ARC, Space Science and
  Astrobiology  Division,  MS 245-3,  Moffett  Field,  CA 94035,  USA}
\email{gdangelo@arc.nasa.gov}        \and       \author{W.       Kley}
\affil{Universit\"at   T\"ubingen,   Institut   f\"ur  Astronomie   \&
  Astrophysik,  Abt. Computational Physics,  Auf der  Morgenstelle 10,
  D-72076  T\"ubingen,  Germany} \email{wilhelm.kley@uni-tuebingen.de}
\altaffiltext{2}{Send    offprint    requests    to    fmasset@cea.fr}
\altaffiltext{3}{NASA Postdoctoral Fellow.}
\begin{abstract}
  The increase  of computational  resources has recently  allowed high
  resolution,  three dimensional calculations  of planets  embedded in
  gaseous protoplanetary  disks. They provide estimates  of the planet
  migration timescale that can  be compared to analytical predictions.
  While  these predictions  can  result in  extremely short  migration
  timescales  for  cores  of  a  few Earth  masses,  recent  numerical
  calculations have given an  unexpected outcome: the torque acting on
  planets  with  masses between  $5\;M_\oplus$  and $20\;M_\oplus$  is
  considerably  smaller  than the  analytic,  linear estimate.   These
  findings  motivated the present  work, which  investigates existence
  and origin of  this discrepancy or ``offset'', as  we shall call it,
  by means  of two and  three dimensional numerical  calculations.  We
  show  that  the  offset  is  indeed physical  and  arises  from  the
  coorbital  corotation torque,  since  (i) it  scales  with the  disk
  vortensity gradient,  (ii) its asymptotic value depends  on the disk
  viscosity, (iii) it is associated to an excess of the horseshoe zone
  width.   We  show  that  the  offset corresponds  to  the  onset  of
  non-linearities  of the  flow  around the  planet,  which alter  the
  streamline topology  as the planet  mass increases: at low  mass the
  flow  non-linearities  are confined  to  the  planet's Bondi  sphere
  whereas at  larger mass the streamlines display  a classical picture
  reminiscent of  the restricted three  body problem, with  a prograde
  circumplanetary disk  inside a ``Roche lobe''.  This  behavior is of
  particular   importance    for   the   sub-critical    solid   cores
  ($M\la15\;M_\oplus$)  in thin  ($H/r\la0.06$)  protoplanetary disks.
  Their migration could be  significantly slowed down, or reversed, in
  disks with shallow surface density profiles.
\end{abstract}
\section{Introduction\label{sec:intro}}
Ever since it was realized that the torque exerted by a protoplanetary
disk onto an orbiting protoplanet  could vary its semi-major axis on a
time  scale much  shorter than  the disk  lifetime  \citep{gt79}, many
efforts have  been made  to determine the  direction and rate  of this
semi-major axis change, referred to as planetary migration. During two
decades,   this  problem  was   essentially  tackled   through  linear
analytical estimates  of the disk torque onto  a point-like perturber.
The  torque on a  planet in  a circular  orbit can  be split  into two
components:  the  differential  Lindblad  torque  and  the  corotation
torque. Early work on planetary migration consisted in determining the
sign   and  value   of   the  differential   Lindblad   torque  in   a
two-dimensional  disk \citep{w86},  since this  torque, in  the linear
regime,  typically   exceeds  the  coorbital   corotation  torque  and
therefore   dictates  the   direction  and   timescale   of  planetary
migration. This work indicated  that planetary migration in most cases
corresponds to an  orbital decay towards the center, and  that it is a
fast process,  thus posing a  threat for the survival  of protoplanets
embedded  in  protoplanetary  disks.   Later efforts  focused  on  the
corotation torque  \citep{w89} and on  the disk's vertical  extent and
pressure effects  on the differential  Lindblad torque \citep{arty93}.
The  analytical  predictions in  the  linear  regime  were checked  by
numerical  integration  of  the differential  equations  \citep{kp93}.
Finally, \citet{tanaka} have given  an expression of the tidal torque,
in the  linear regime, that takes  into account both  the Lindblad and
coorbital corotation  torques, and that  fully takes into  account the
three  dimensional  structure  of   the  disk.   These  analytical  or
semi-analytical studies  all consider small mass planets,  for which a
linear  approximation of the  disk response  is valid.   Other studies
dealt with a strongly non-linear  case, that of embedded giant planets
\citep{lp86a,lp86b}. They showed that a giant planet tidally truncates
the disk by opening a gap around its orbit, and that it is then locked
in the viscous disk evolution,  a process that was much later referred
to as type II migration  \citep{w97}.  A more recent work \citep{mp03}
considers the case of sub-giant  planets (planets which have a mass of
the order  of a Saturn mass, if  the central object has  a solar mass)
embedded  in  massive  disks.   This  work shows  that  the  coorbital
corotation torque may  have a strong impact on  the migration, and can
lead to a  runaway of the latter, either inwards  or outwards. As this
mechanism  heavily  relies upon  the  finite  width  of the  horseshoe
region, it also  corresponds to a non-linear mechanism.   The onset of
non-linear  effects should  therefore occur  below a  sub-giant planet
mass, but the first manifestation of these effects and their impact on
planetary   migration   have   not   been   investigated   thus   far.
\citet{kpap96}, by writing the  flow equations in dimensionless units,
have shown  that the flow  non-linearity is controlled by  a parameter
${\cal  M}=q^{1/3}/h$, where $q=M_p/M_*$  is the  planet mass  to star
mass  ratio and $h=H/r$  is the  disk aspect  ratio. The  linear limit
corresponds  to ${\cal  M}\rightarrow 0$,  while the  condition ${\cal
  M}>1$  has  been  considered   as  a  necessary  condition  for  gap
clearance,  and has  sometimes been  referred  to as  the gap  opening
thermal  criterion,  although a  recent  work  by \citet{crida06}  has
revisited the conditions for gap opening.

In the  last few  years, the increase  of computational  resources has
made possible the evaluation of the disk torque exerted on an embedded
planet by means of hydrodynamical calculations, both in two dimensions
\citep{las99,n00,dhk02,m02,nb03a,nb03b}     and    three    dimensions
\citep{dkh03,bate03},  both  for  small  mass planets  and  for  giant
planets.   In  particular,  the  case  of small  mass  planets  allows
comparison  with  analytical  linear  estimates.   This  was  done  by
\citet{dhk02,dkh03} and \citet{bate03},  who compared the torques they
measured  with the estimate  by \cite{tanaka}.  Although \citet{dhk02}
and  \citet{bate03}  found  results  in  good  agreement  with  linear
expectations, \citet{dkh03} found a significant discrepancy for planet
masses  in  the range  $5$--$20\;M_\oplus$.  Namely,  they found  that
migration in  this planet  mass range  may be more  than one  order of
magnitude  slower than expected  from linear  estimates.  In  the same
vein, \citet{m02}  found that planetary migration for  the same planet
masses  can be  much  slower,  or even  reversed,  compared to  linear
estimates.  Since  the migration of protoplanetary cores  of this mass
constitutes a bottleneck  for the build up of  giant planets cores (as
this build up is slow, while the migration of these cores is fast), it
is  fundamental to  establish  whether  this effect  is  real and,  if
confirmed,  to investigate  the reasons  of this  behavior.   We shall
hereafter refer to this discrepancy as {\em the offset}.

We adopt for the presentation of our results a heuristic approach that
consists first in presenting the set of properties that we could infer
from our calculations, and  then in interpreting and illustrating them
through the  appropriate analysis.  Besides  its pedagogical interest,
this approach also closely follows our own approach to this problem.

In section~\ref{sec:setup}, we describe the two independent codes that
we  used to  check  the properties  of  the offset,  and  we give  the
numerical    setup    used   by    each    of    these   codes.     In
section~\ref{sec:properties}  we list  the  set of  properties of  the
offset that our numerical experiments allowed us to identify, namely:
\begin{itemize}
\item The  offset scales with the vortensity  gradient (the vortensity
  being defined as the vertical  component of the vorticity divided by
  the surface density).
\item The offset value  varies over the horseshoe libration timescale,
  and tends  to small  values at small  viscosity, whereas  it remains
  large at high viscosity.
\item The maximum relative offset occurs for a planet mass that scales
  as $h^3$.
\end{itemize}
We then interpret these properties  as due to a non-linear behavior of
the coorbital  corotation torque that exceeds  its linearly estimated
value.   Using  the  link  between  coorbital  corotation  torque  and
horseshoe zone drag  \citep{wlpi91,wlpi92,m01,m02,mp03}, we perform in
section~\ref{sec:stream}  a  streamline  analysis  in order  to  check
whether  the coorbital  corotation torque  excess is  associated  to a
horseshoe zone  width excess.  We find  that this is  indeed the case.
In  section~\ref{sec:bernou},  we  relate  this width  excess  of  the
horseshoe region to a transition  of the flow properties in the planet
vicinity, from  the linear regime  to the large  mass case in  which a
circumplanetary  disk surrounds  the  planet.  We  finally discuss  in
section~\ref{sec:discuss} the  importance of these  properties for the
migration  of sub-critical  solid cores.   We  sum up  our results  in
section~\ref{sec:conclusion}.

\section{Hydrodynamical codes and numerical set up}
\label{sec:codes}
We  used  two independent  hydro-codes  to  perform  our tidal  torque
estimates.  One of these codes is the 3D nested grid code NIRVANA, the
other one  is the  2D polar code  FARGO.  The  use of these  codes was
complementary:  while FARGO suffers  from the  2D restriction  and its
outcome is plagued by the  use of a gravitational softening length, it
enables  one to  perform a  wide  exploration of  the parameter  space
(mainly, in our  case, in term of planet  mass, surface density slope,
disk thickness  and viscosity). The properties suggested  by the FARGO
runs can  later be confirmed by  much more CPU-demanding  3D runs with
NIRVANA.
\subsection{The NIRVANA code}
This code is a descendant of  an early version of the MHD code NIRVANA
\citep{ziegler1997}, hence the name.  For the current application, the
magnetic terms in the MHD  equations are excluded. The code features a
covariant  Eulerian  formalism  that  allows  to  work  in  Cartesian,
cylindrical,  or spherical  polar coordinates  in one,  two,  or three
dimensions. The MHD  equations are solved on a  staggered mesh, with a
constant  spacing  in  each  coordinate direction  via  a  directional
splitting procedure,  whereby the advection part and  the source terms
are dealt with separately. The advection of the hydrodynamic variables
is performed  by means of a  second-order accurate scheme  that uses a
monotonic slope limiter  \citep{vl77}, enforcing global conservation
of mass  and angular  momentum.  Viscous forces  are implemented  in a
covariant tensor formalism.  The  code allows a static mesh refinement
through a hierarchical  nested-grid structure \citep{dhk02,dkh03}. The
resolution increases by a factor $2$ in each direction from a sub-grid
level to the  next nested level. When employed in  a 3D geometry, this
technique produces  an effective refinement  of a factor $8$  from one
grid level to the next one.
\subsection{The FARGO code}
The FARGO  code is a staggered  mesh hydro-code on a  polar grid, with
upwind  transport   and  a   harmonic,  second  order   slope  limiter
\citep{vl77}.   It solves the  Navier-Stokes and  continuity equations
for a Keplerian disk subject to  the gravity of the central object and
that of embedded protoplanets.  It  uses a change of rotating frame on
each ring  that enables  one to increase  significantly the  time step
\citep{m00,m00b}.   The hydrodynamical solver  of FARGO  resembles the
widely  known  one of  the  ZEUS  code  \citep{zeus}, except  for  the
handling of momenta advection.  The Coriolis force is treated so as to
enforce  angular  momentum  conservation  \citep{k98}.   The  mesh  is
centered  on the  primary. It  is therefore  non-inertial.   The frame
acceleration  is  incorporated   in  a  so-called  potential  indirect
term. The full viscous stress tensor in cylindrical coordinates of the
Navier-Stokes equations is implemented in FARGO.  A more detailed list
of  its properties  can  be found  on  its website\footnote{See:  {\tt
http://www.maths.qmul.ac.uk/$\sim$masset/fargo}}.

\subsection{Units}

As  is  customary  in  numerical  calculations  of  disk-planet  tidal
interactions, we use the planet orbital radius $a$ as the length unit,
the  mass  of  the  central   object  $M_*$  as  the  mass  unit,  and
$(a^3/GM_*)^{1/2}$ as  the time unit,  where $G$ is  the gravitational
constant,  which is $G=1$  in our  unit system.   Whenever we  quote a
planet mass  in Earth masses, we  assume the central object  to have a
solar mass. We  note $M_p$ the planet mass  and $q=M_p/M_*$ the planet
to star mass ratio.

\subsection{Numerical Set up}
\label{sec:setup}
Both codes  use an  isothermal equation of  state with a  given radial
temperature  (or sound  speed)  profile.  If  $P$  is the  (vertically
integrated for  FARGO) pressure and $\rho$  the (vertically integrated
for FARGO) gas  density, then the equation of  state is $P=c_s^2\rho$.
The  disk  vertical  scale-height  is  $H(r)=c_s(r)/\Omega(r)$,  where
$\Omega(r)$  is the disk  angular frequency  at radius~$r$.   The disk
aspect ratio,  $h(r)=H(r)/r$, is  taken uniform in  the disks  that we
simulate,  and  it  varies  from  $h=  0.03$  to  $h=  0.06$
depending on the runs.

The  softening  length is  applied  to  the  planet potential  in  the
following manner:

\begin{equation}
\Phi_p=-\frac{GM_p}{\sqrt{r_p^2+\epsilon^2}},
\end{equation}

where  $\Phi_p$ is  the planet  potential,  $r_p$ the  distance to  the
planet, and $\epsilon$ is the softening length.

In all the runs presented in this  work, the planet is held on a fixed
circular  orbit.   Moreover,  there  is  no  gas  accretion  onto  the
planet. This is quite different from the prescription of \citet{dkh03}
and  \citet{bate03}.   However,  we  shall  see  that  the  effect  we
investigate  is  related to  the  coorbital  corotation torque,  which
itself is  related to  the horseshoe dynamics.   In the case  in which
accretion is allowed, the flow topology in the planet vicinity is more
complex than in a non-accreting  case, with an impact on the horseshoe
zone and on the coorbital  corotation torque value. In order to retain
only  the physics  relevant to  the effect  we are  interested  in, we
discard gas accretion  onto the planet.  It should  however be kept in
mind    that   this    is    not   realistic    for   planet    masses
$M_p\ga15\;M_\oplus$.   Nonetheless, the  phenomenon we  describe does
persist, and indeed was  originally observed, when planetary cores are
allowed to accrete.

In our runs the disk surface density is initially axisymmetric and has
a  power-law profile:  $\Sigma(r) =  \Sigma_0(r/r_0)^{-\alpha}$, where
$r_0=1$ is the radius at which the surface density is $\Sigma_0$.  The
kinematic  viscosity has  a  uniform  value over  the  disk.  We  have
adopted  a   reference  set  up   which  closely  resembles   the  one
of~\citet{dkh03} or \citet{bate03}.  Its characteristics are listed in
Table~\ref{tab:reference}.   Whenever  we   vary  one  disk  parameter
(e.g. aspect  ratio or viscosity),  we adopt for the  other parameters
the reference values.  For a  given set of disk parameters, we perform
several calculations with different planet masses.
\begin{deluxetable}{lcc}
\tablewidth{0pt} \tablecaption{Disk parameters for the reference
calculations. There is no accretion onto the planets.\label{tab:reference}}
\tablehead{\colhead{Parameter} & \colhead{Notation} & \colhead{Reference value}} 
\startdata 
Aspect ratio                  & $h$          & $0.05$\\
Surface density slope         & $\alpha$     & $1/2$\\
Viscosity                     & $\nu$        & $10^{-5}$\\
\enddata
\end{deluxetable}

We list below the details specific to each code:
\begin{itemize}
\item In the 3D-NIRVANA runs,  the computational domain is a spherical
  sector $[R_{\mathrm{min}},R_{\mathrm{max}}]\times%
  [\theta_{\mathrm{min}},\theta_{\mathrm{max}}]\times2\pi$,       whose
  radial         boundaries         are        $R_{\mathrm{min}}=0.4$,
  $R_{\mathrm{max}}=2.5$.   Symmetry is assumed  relative to  the disk
  mid-plane and  only the upper half  of the disk  is simulated, hence
  $\theta_{\mathrm{max}}=90^{\circ}$.     The   minimum   co-latitude,
  $\theta_{\mathrm{min}}$, varies from $80^{\circ}$ to $82.5^{\circ}$,
  according to the value of  the aspect ratio $h$. The vertical extent
  of  the   disk  comprises  at  least   $3$  pressure  scale-heights.
  Outgoing-wave  (or non-reflecting) boundary  conditions are  used at
  the inner radial border \citep{g96}.  In order to exploit the mirror
  symmetry of  the problem with  respect to the disk  equatorial plane
  (the disk  and the planet  orbit are coplanar), a  symmetry boundary
  condition  was used  at  the  disk mid-plane,  which  enables us  to
  simulate  only the  upper  half of  the  disk.  Finally,  reflecting
  boundary conditions were  used at the outer radial  border, which is
  located sufficiently far from the  orbit so that the wake reflection
  will     not    alter    our     torque    evaluation,     and    at
  $\theta=\theta_{\mathrm{min}}$, were the  matter is so rarefied that
  the choice of the boundary  condition has virtually no impact on the
  flow properties  on the bulk of  the disk.  The  reference frame has
  its  origin on  the center  of mass  of the  star-planet  system and
  corotates with the  planet.  The grid hierarchy consists  of a basic
  mesh with  $(N_{R},N_{\theta},N_{\phi})=(143,13,423)$ grid zones and
  $4$ additional  sub-grid levels  centered at the  planet's position,
  each  with $(64,12,64)$  grid zones.   The initial  vertical density
  distribution  is   that  of  an  unperturbed   disk  in  hydrostatic
  equilibrium,     which     in     spherical    coordinates     reads
  $\rho(t=0)=\rho_{0}(R)\,\exp{[(\sin{\theta}-1)/h^2]}/%
  (\sin{\theta})^{(\alpha+1)}$,         where         $\rho_{0}\propto
  1/R^{(\alpha+1)}$  and  the  sound  speed  is assumed  to  scale  as
  $c_s\propto   h/\sqrt{R\sin{\theta}}$.   Such  density   profile  is
  stationary in  the limit $M_{p}\rightarrow 0$.   The initial surface
  density, obtained  by integrating the  mass density in  $\theta$, is
  $\Sigma=\Sigma_{0}(a/R)^{\alpha}$,    where    $\Sigma_{0}=2.9\times
  10^{-4}$.  Calculations were performed for many values of the planet
  to star mass ratio, from $q=10^{-6}$ to $q=2\times10^{-4}$, in disks
  with various  values of the  initial density slope  $\alpha$, aspect
  ratio $h$, and kinematic  viscosity $\nu$.  Simulations were run for
  up  to  $140$  orbital  periods  to  measure  (partially)  saturated
  coorbital corotation  torques. Shorter runs ($10$  orbits) were used
  to  monitor  (partially)  unsaturated  corotation  torques.   In  3D
  calculations,  torques arising  from  the Roche  lobe  are not  very
  sensitive to  the choice of the softening  parameter, $\epsilon$, in
  the  planet  gravitational potential,  as  long  as  it is  a  small
  fraction   of   the  Hill   radius   $R_H=a(q/3)^{1/3}$.   We   used
  $\epsilon=0.1R_{H}$.   However, some  models  were also  run with  a
  smaller softening length and produced no significant differences.
\item  In the 2D-FARGO  runs, the  mesh inner  boundary is  at $R_{\rm
    min}=0.5$ and the  mesh outer boundary is at  $R_{\rm max}=2.1$. A
  non-reflecting boundary  condition was  used at each  boundary.  The
  resolution is of $N_{\rm  rad}=153$ zones in radius and $N_\phi=600$
  zones in azimuth.  The mesh spacing is uniform both in radius and in
  azimuth.   The  frame  corotates  with  the planet.   The  value  of
  $\Sigma_0$ is  $6\cdot 10^{-4}$.  The potential  softening length is
  $\epsilon=0.3H$.  This value is quite low.  Preliminary calculations
  have  shown  that the  offset  is  much  larger at  small  potential
  softening length  value, which is why  we adopted this  value. For a
  given set of  disk parameters, we performed 35  calculations with 35
  different  planet masses,  in  geometric sequence  and ranging  from
  $q=10^{-6}$  to $q=10^{-3.5}$:  $q_i=10^{-6+2.5i/34}$,  $0\leq i\leq
  34$.  Most of the calculations are  run for 100 orbits, in order for
  the coorbital  corotation torque to saturate if  the disk parameters
  imply its saturation.   We have also performed series  of short runs
  for 10 orbits, in order to have an unsaturated corotation torque.
\end{itemize}

\subsection{Torque evaluation}
\begin{itemize}
\item In the  3D runs, the gravitational torques  acting on the planet
  are  evaluated either  every  $5$ orbits  (long-run simulations)  or
  every orbit  (short-run simulations). In  the first case,  the total
  torque  is  averaged over  the  last  $30$  orbital periods  of  the
  calculation whereas, in  the second case, it is  averaged from $t=7$
  and  $t=10$   orbits.   As  mentioned   in  section~\ref{sec:setup},
  accretion onto  the planetary core is  not allowed. In  the low mass
  limit,  this leads to  the formation  of a  gas envelope  around the
  planet. The size  of the envelope depends on the core  mass and is a
  fraction  of   $R_{H}$.   To  avoid  the   envelope  region,  torque
  contributions  from  within the  Hill  sphere  were discarded.  This
  choice  may occasionally  result  in some  corotation torques  being
  unaccounted for.   When this happens, the departure  from the linear
  regime may be underestimated.  However, tests performed by excluding
  torques from  a region of radius  $0.5R_{H}$\footnote{The net torque
    exerted by material deep inside the Hill sphere of a non-accreting
    planet  is  negligible  if  density  gradients  are  appropriately
    resolved  \citep{dbl05}.}   indicate that  the  effects would  not
  significantly  change the  results  of this  study. Therefore,  this
  choice is  conservative since it may  occasionally underestimate the
  excess of coorbital corotation torques but assures that our analysis
  is not affected by spurious  torques from material possibly bound to
  the core.
\item In  FARGO, the  torque exerted  by the disk  onto the  planet is
  evaluated every $1/20^{th}$  of orbit. In the long  runs case ($100$
  orbits), the torque value is averaged from $t=40$ to $t=100$~orbits,
  in order to  discard any transient behavior at  the beginning of the
  calculation, due to  corotation torque (possibly partial) saturation
  on the  libration timescale.  In  the short runs case,  we generally
  take (unless otherwise stated)  the torque average between $t=6$ and
  $t=7$~orbits.  We also entertained  the issue whether the Roche lobe
  material must be taken into account. The FARGO code, in its standard
  version, outputs both the torque exerted by the totality of the disk
  onto  the planet,  without a  special  treatment of  the Roche  lobe
  material,  and the torque  obtained by  tapering the  torque arising
  from the Roche lobe  and its surroundings by $1-\exp[-(r_p/R_H)^2]$,
  where $r_p$ is  the zone center distance to the  planet.  We show in
  section~\ref{sec:properties}  that  taking  or  not the  Roche  lobe
  content into account does  make a difference, but that qualitatively
  one obtains the offset properties  in both cases.  We have chosen to
  include  the Roche  lobe content  in the  torque evaluation  for the
  FARGO calculations presented in  this work.  There is another reason
  for this choice,  namely that the material that  should be discarded
  in  the torque  calculation  should  be the  one  pertaining to  the
  circumplanetary  disk: one would  define the  system of  interest as
  \{the planet  + the circumplanetary disk\}.   If the circumplanetary
  disk has a radius that scales with $R_H$ and that amounts to several
  $0.1R_H$ for  large planet  masses, this is  not true for  the small
  planet masses that represent most of the mass interval over which we
  perform   the   calculations.    For   these   small   masses,   the
  circumplanetary disk has a radius  much smaller than a few $0.1R_H$,
  or may not even exist, as we shall see in section~\ref{sec:bernou}.
\end{itemize}

\section{Offset properties}
\label{sec:properties}
\subsection{Reference run}
\subsubsection{2D results}
\begin{figure}
\plotone{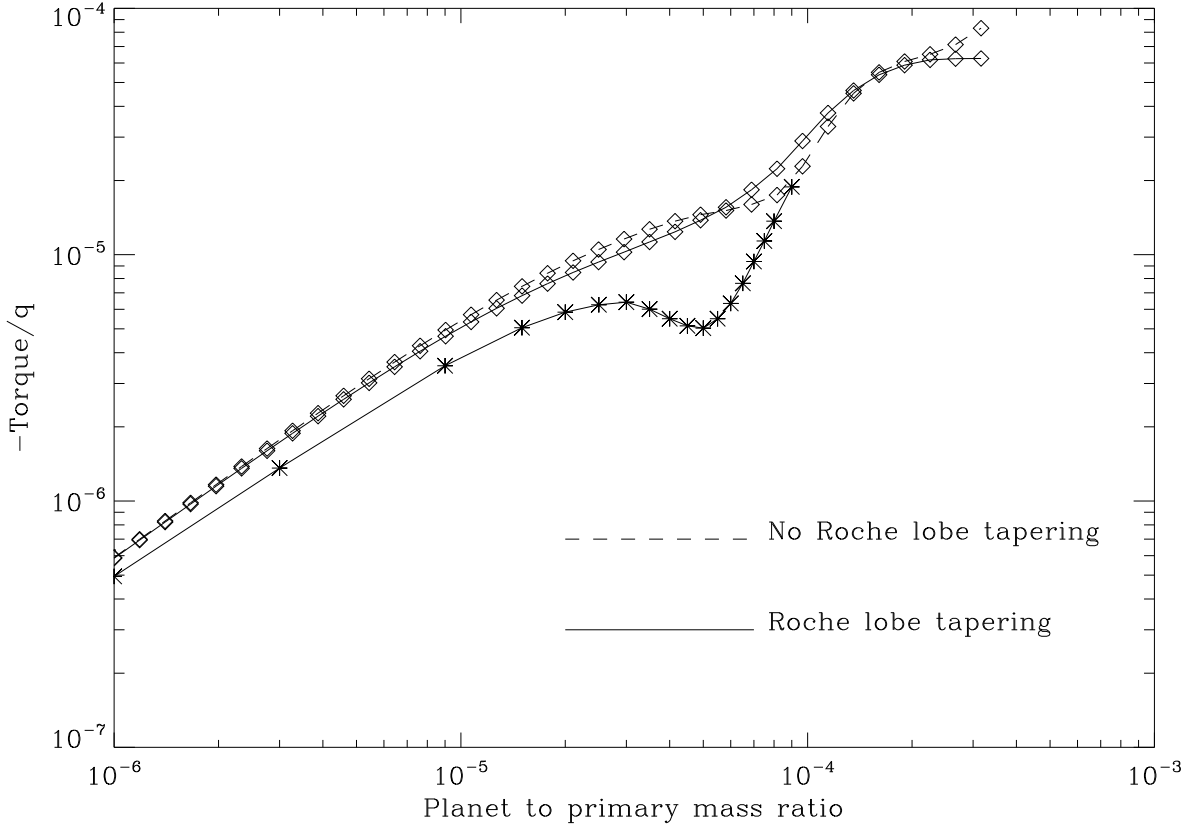}
\caption{\label{fig:ref2d3d}Negative  specific  torque  acting on  the
  planet, as a function of its  mass, in the reference disk for the 2D
  case. The  solid line with  diamonds shows the torque  computed with
  Roche lobe tapering, while the dashed line shows the torque computed
  without special  treatment of the  Roche lobe zones. The  solid line
  with stars shows the  results of the three dimensional calculations,
  scaled  by  $\Sigma_0^{2D}/\Sigma_0^{3D}=60/29$.  We note  that  the
  offset depth is larger in the 3D case.}
\end{figure}

Fig.~\ref{fig:ref2d3d}  shows  the   results  of  the  reference  run,
corresponding  to the  parameters  of Table~\ref{tab:reference},  both
with and without Roche lobe tapering. Both curves show the offset near
$q=10^{-4}$. However the  curves do not coincide, and the offsets have
slightly different shapes, which indicates  that it is due to material
located inside of the Hill sphere or in its immediate vicinity.

As mentioned in the Introduction, previous two-dimensional simulations
by \citet{dhk02}  have apparently missed  the offset feature  shown in
Figure~\ref{fig:ref2d3d}.  The most likely reason why this happened is
the use  of an extremely small  softening parameters (on  the order of
$0.02R_H$),  associated with  the action  of torques  deep  inside the
planet's  Hill  sphere  (at  distances from  the  planet  $r_p\lesssim
0.2R_H$).  We  shall see in  section~\ref{sec:bernou} that for  such a
small softening length we should  expect the offset feature to peak at
$q< 10^{-6}$, which is not in the mass range covered by \citet{dhk02}.
Furthermore,  their  analysis  is  complicated  by  the  inclusion  of
accretion and  the presence  of a  gap or dip  in the  initial surface
density profile.  We also performed a set of calculations with NIRVANA
in  2D mode,  using  the  reference parameters  and  adopting a  setup
similar to  that of FARGO.   The resulting specific torque  versus the
planetary  mass is  consistent with  the solid  line with  diamonds in
Figure~\ref{fig:ref2d3d}. 
\subsubsection{3D results}
The behavior of  the total specific torque exerted by  the planet on a
three-dimensional    disk,    for     the    parameters    given    in
Table~\ref{tab:reference}, is illustrated in Figure~\ref{fig:ref2d3d}.
The departure from the total  torque predicted by the linear theory is
largest     at    $q=5\times10^{-5}$.      A     comparison    between
Figure~\ref{fig:ref2d3d}  and  Figure~6  in  \citet{dkh03}  allows  to
evaluate  the impact  of core  accretion on  the excess  of corotation
torques.    This   represents   an   important  issue   since   around
$10\;M_{\oplus}$  the runaway gas  accretion phase  is most  likely to
occur  \cite[e.g.,][]{kw93,petal96,hbl05}.   Accretion  on the  planet
seems to enhance the excess  of coorbital corotation torques, over the
predictions based on the linear  regime, since it affects the width of
the  horseshoe  region.  The  location  where  the  offset is  maximal
recedes  from  $q=5\times10^{-5}$, when  cores  are non-accreting,  to
$q=3\times10^{-5}$, when cores accrete at maximum rate.

\subsection{Dependence on the vortensity gradient}
\subsubsection{2D results}
\begin{figure}
\plottwo{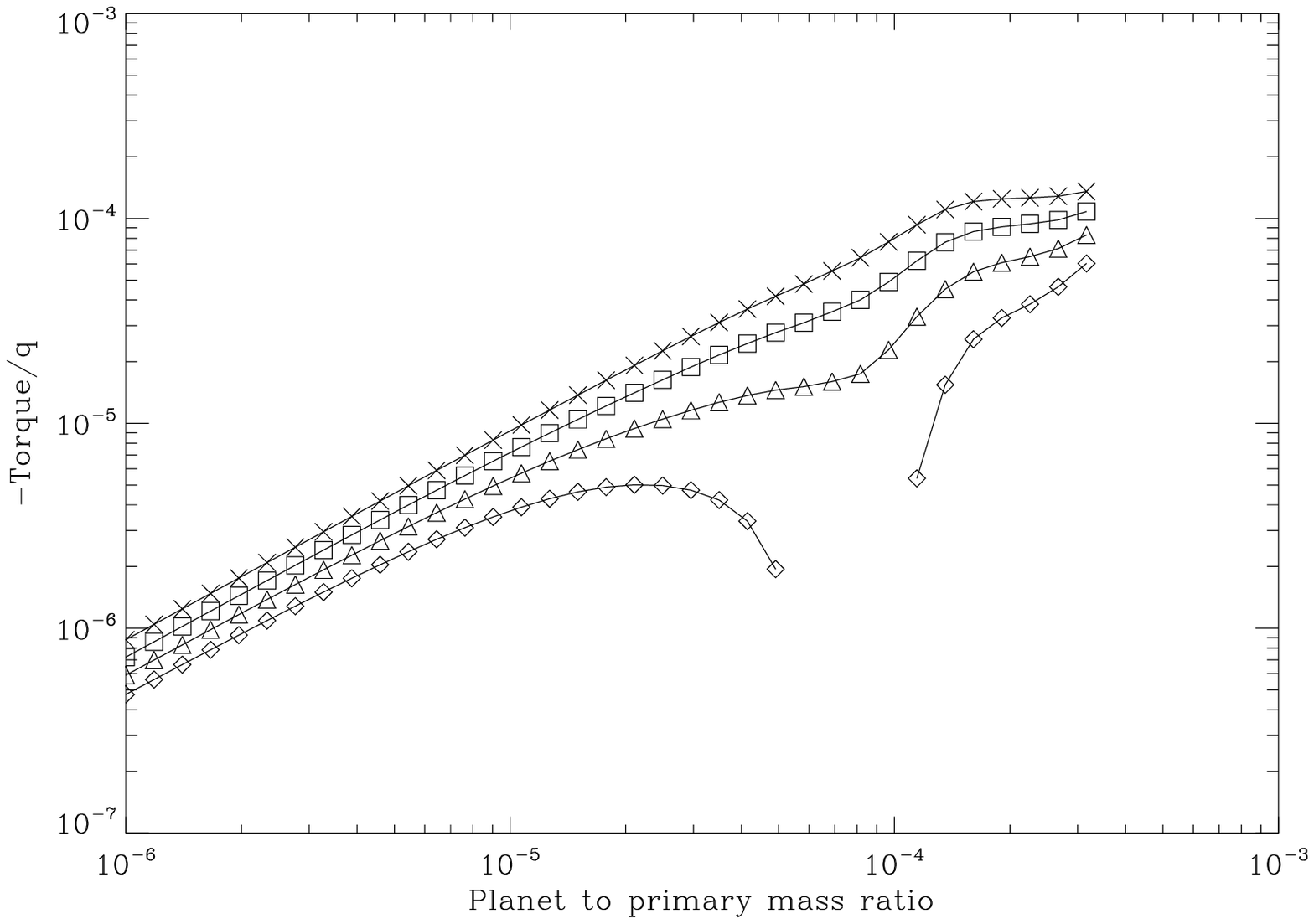}{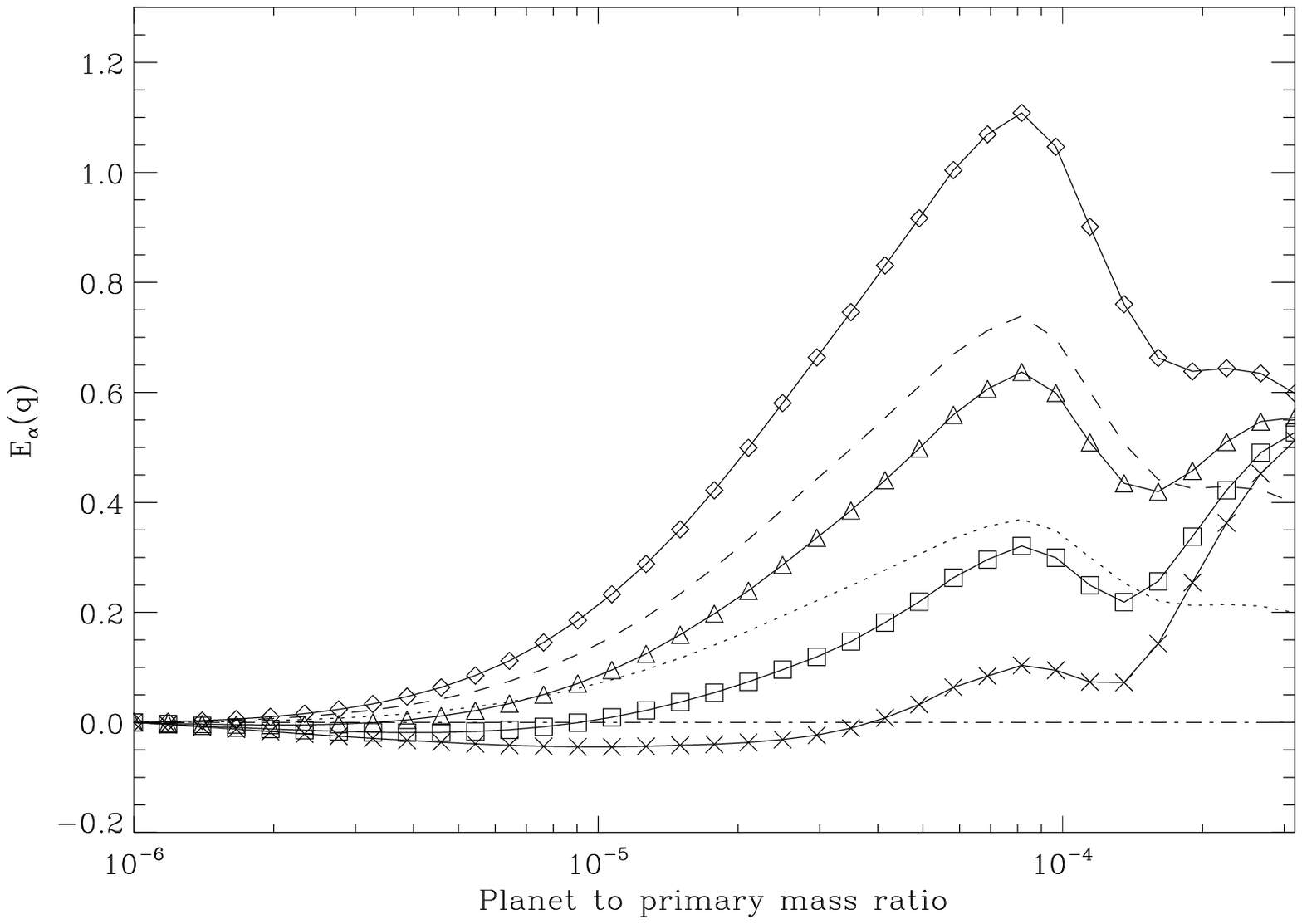}
\caption{\label{fig:depbetafargo} 2D results.
  Left:  Negative  of  specific  torque
  acting on the planet, as a  function of its mass, for four values of
  the  surface  density  slope:  $\alpha=0$  (diamonds),  $\alpha=1/2$
  (reference   calculation,  triangles),   $\alpha=1$   (squares)  and
  $\alpha=3/2$ (crosses).  The hole in  the data for the  flat surface
  density profile  corresponds to a torque  reversal. Right: departure
  from linearity for the same surface density slopes (same symbols).
  The meaning of the additional lines is explained in the text.}
\end{figure}

Fig.~\ref{fig:depbetafargo}a   shows   the  results   of   a  set   of
calculations  with  four  different  disks, having  different  surface
density slopes.   The set  that exhibits the  smallest departure  to a
linear trend  (straight line) corresponds  to $\alpha=3/2$, i.e.  to a
flat  vortensity  profile,  since  $d  \log
(\Sigma/B)/dr=3/2-\alpha$.

Fig.~\ref{fig:depbetafargo}b shows the quantity
\begin{equation}
\label{eqn:ealpha}
E_\alpha(q)=1-\frac{T_\alpha(q)}{q^2}
\frac{q_{\rm min}^2}{T_\alpha(q_{\rm min})},
\end{equation}
where $T_\alpha(q)$  is the disk torque  on the planet  with planet to
star mass ratio  $q$ when the disk surface  density slope is $\alpha$,
and  $q_{\rm min}$  is  the minimal  mass  ratio in  our sample  (here
$q_{\rm  min}=10^{-6}$). Whenever  the  disk response  is linear,  the
torque  scales  with  $q^2$  and $E_\alpha$  vanishes.   The  quantity
$E_\alpha(q)$ is therefore a measure of the departure from linearity%
\footnote{By  this  we  mean  the  departure  from  the  torque  value
  predicted  by  a linear  analysis  of  the disk-planet  interaction.
  Naturally, it is  also the departure from the  linear scaling of the
  torque with $q^2$.}  of the torque.  It reaches unity when the total
torque cancels out, and exceeds  one when migration is reversed.  From
Fig.~\ref{fig:depbetafargo}b  we  can see  that  for $q>10^{-4}$,  the
torque value differs from  its linearly extrapolated value, regardless
of the vortensity  slope.  For smaller masses, the  departure from the
linearly  predicted  value is  larger  for  larger vortensity  slopes.
Although the  flat surface vortensity profile  ($\alpha=3/2$) does not
have  a vanishing  $E_\alpha$,  it is  nevertheless  the profile  that
exhibits the smallest departure to linear prediction (by at most 10~\%
up to $q\sim  1.5\cdot 10^{-4}$). The dashed and  dotted line show the
curve  of  $E_0$  for   the  flat  surface  density  profile  (maximal
vortensity slope) respectively scaled by $2/3$ and $1/3$. These curves
show  that the departure  to linearity  approximately scales  with the
vortensity slope.
\subsubsection{3D results} 
\begin{figure}
\plottwo{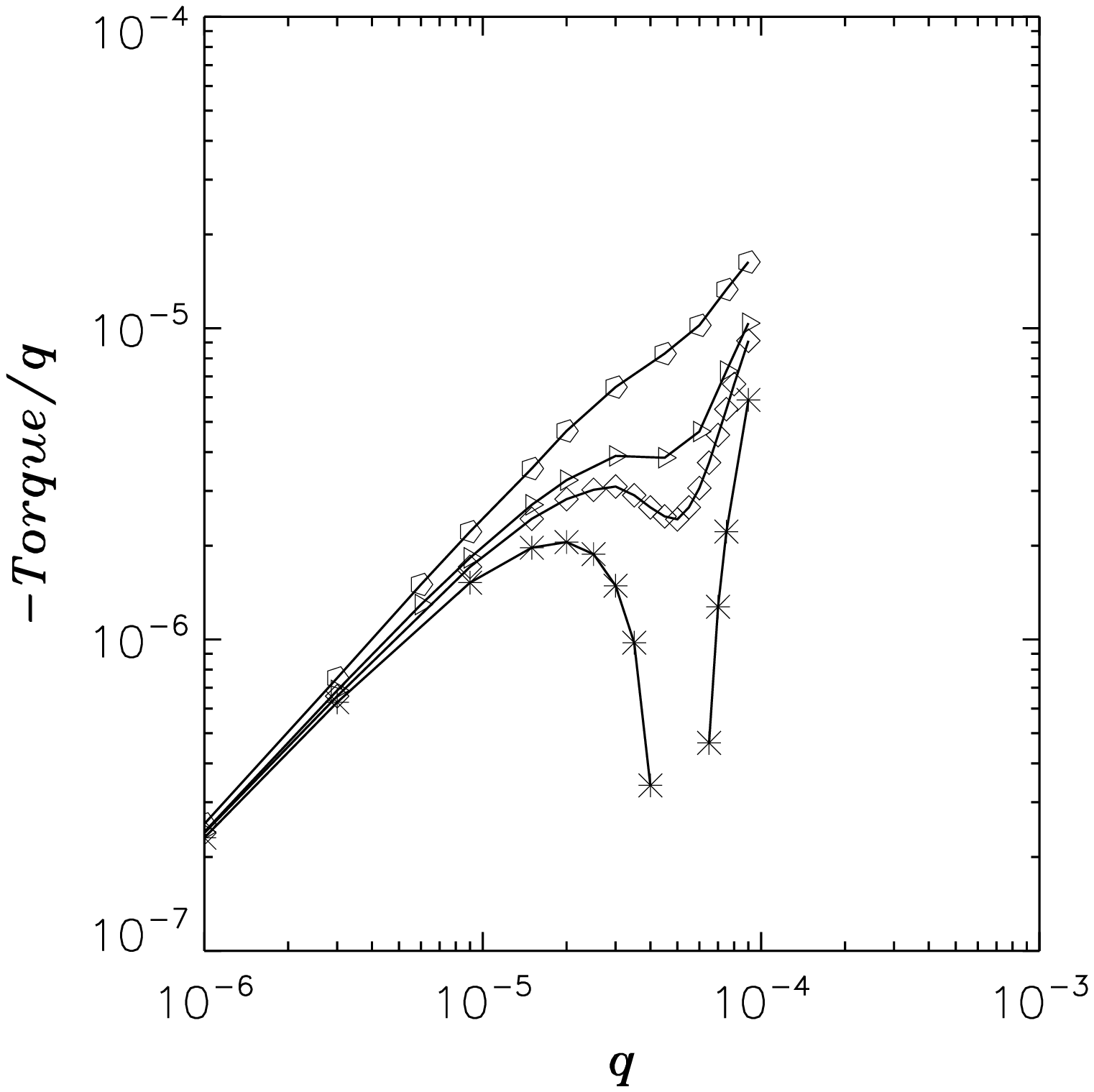}{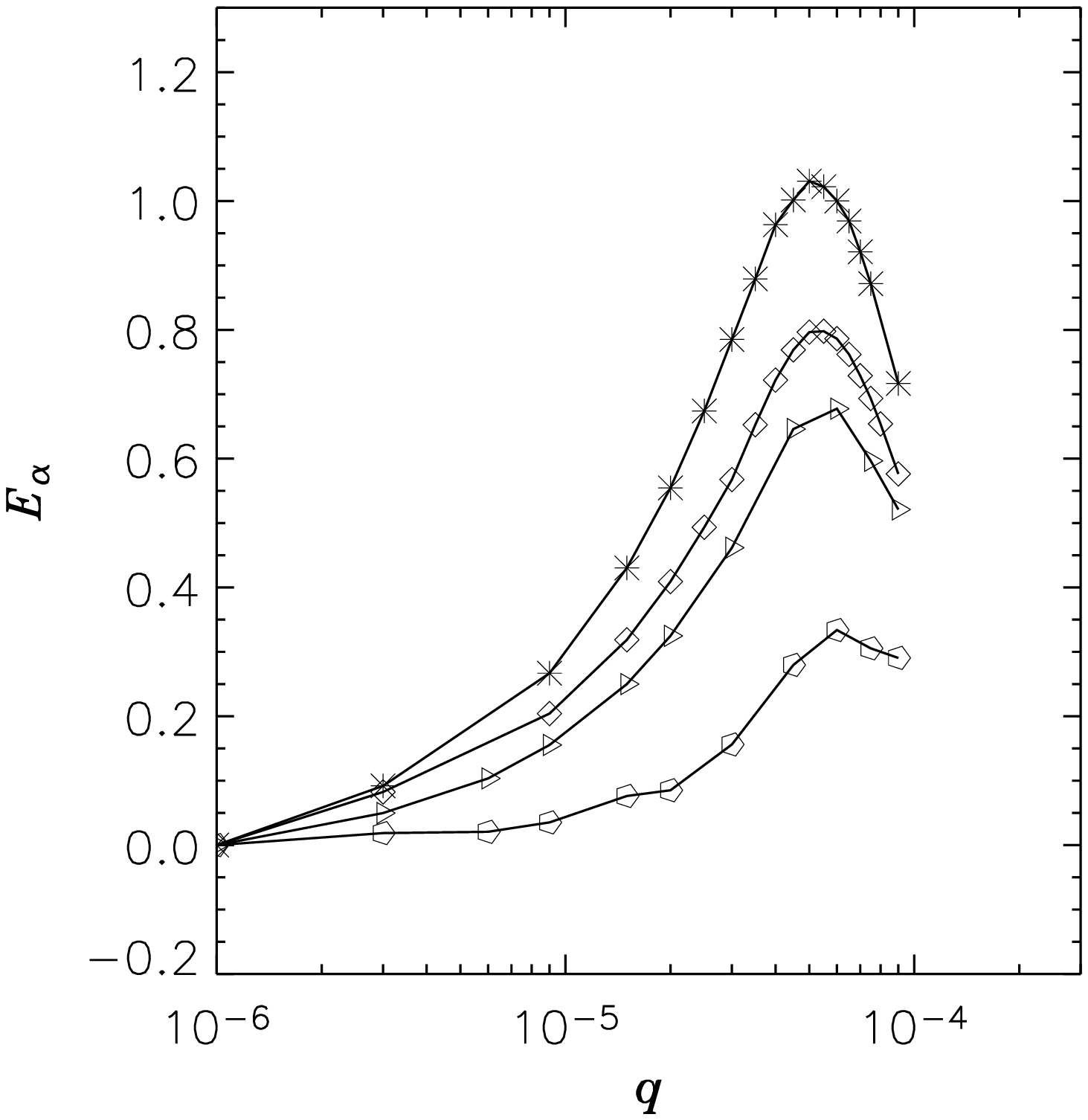}
\caption{\label{fig:depbeta3d} 3D results. 
  Left.  Negative  of the specific torque
  exerted  on the planet,  as a  function of  its mass,  for different
  values  of  the   surface  density  slope:  $\alpha=0$  (asterisks),
  $\alpha=1/2$ (diamonds),  $\alpha=3/4$ (triangles), and $\alpha=3/2$
  (pentagons).   The   gap  in  the  data  for   the  $\alpha=0$  case
  corresponds  to  situations  where  the total  torque  is  positive.
  Right. Departure from  linearity (Eq.~\ref{eqn:ealpha}) for the same
  values of $\alpha$ (same symbols identify same models).}
\end{figure}
The  left panel  of  Figure~\ref{fig:depbeta3d} shows  the
specific torque  exerted by the disk  on the planet,  obtained from 3D
calculations with different  surface density slopes, $\alpha$. Torques
are (partially)  saturated, which means  that they have  reached their
steady state value, which is a fraction of their initial (unsaturated)
value.  The  behavior of the  quantity $E_{\alpha}$ is  illustrated in
right panel for  the same models.  As observed in  the 2D results, the
departure from  the linear (type~I) regime,  increases with increasing
vortensity gradient.

\subsection{Dependence on the viscosity}
\label{sec:viscdep}
The  previous  section suggests  that  the  offset  is linked  to  the
coorbital  corotation  torque, since  it  scales  with the  vortensity
gradient across  the orbit. For the vortensity  slopes considered, the
coorbital  corotation torque acting  on the  planet is  positive. Note
that  as the  offset  corresponds to  a  positive value  added to  the
linearly  expected torque value,  this would  suggest that  the offset
corresponds to a corotation torque larger than predicted by the linear
analysis. If the  offset is indeed due to  the corotation torque, then
it should  depend on the  disk viscosity, since the  corotation torque
depends  on it  \citep{wlpi92,m01,m02,bk01,og03}.  We  have undertaken
additional sets of calculations, in which we take the reference values
of   Table~\ref{tab:reference},   except  that   we   vary  the   disk
viscosity~$\nu$.
\subsubsection{2D results}
\begin{figure}
\plottwo{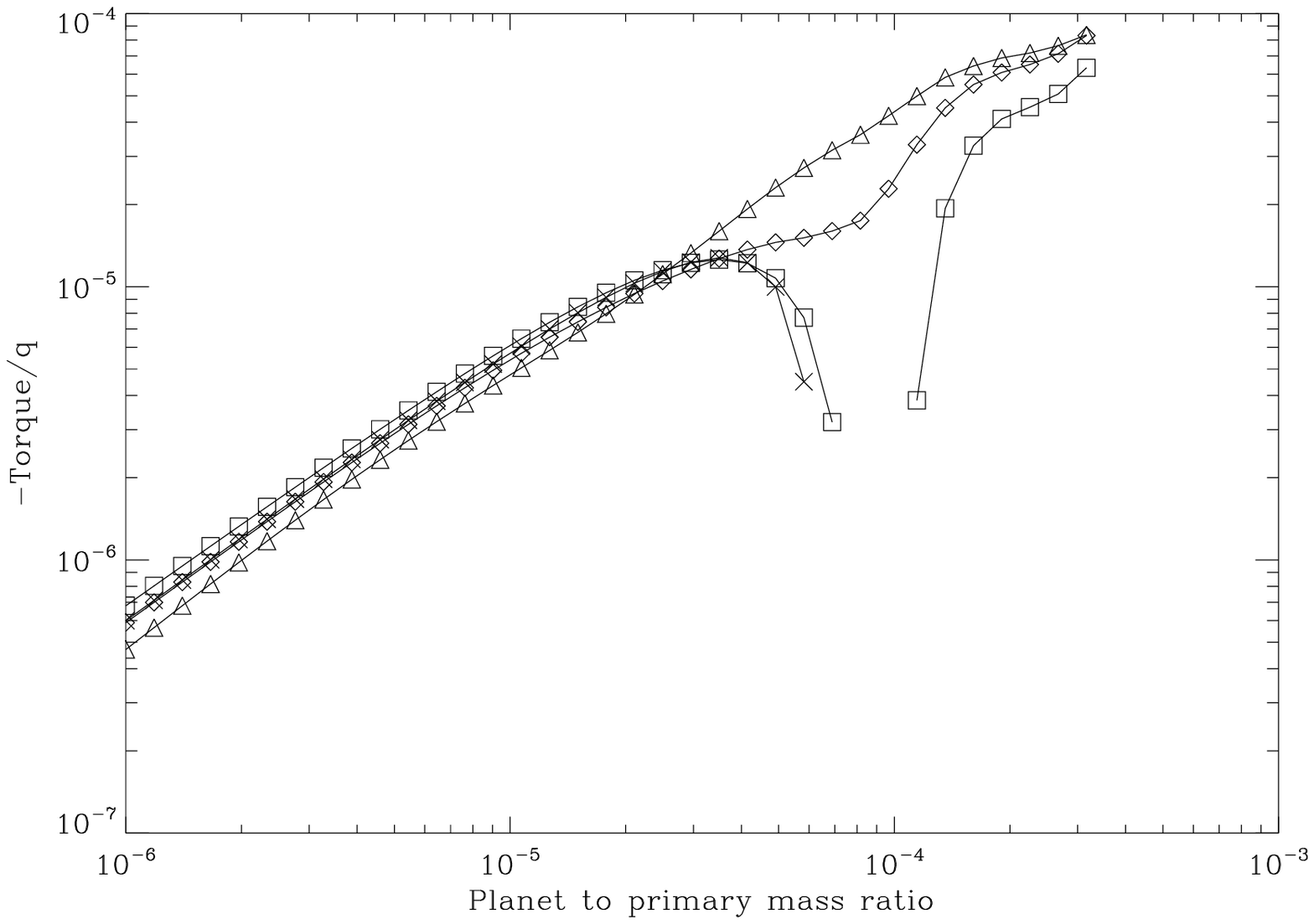}{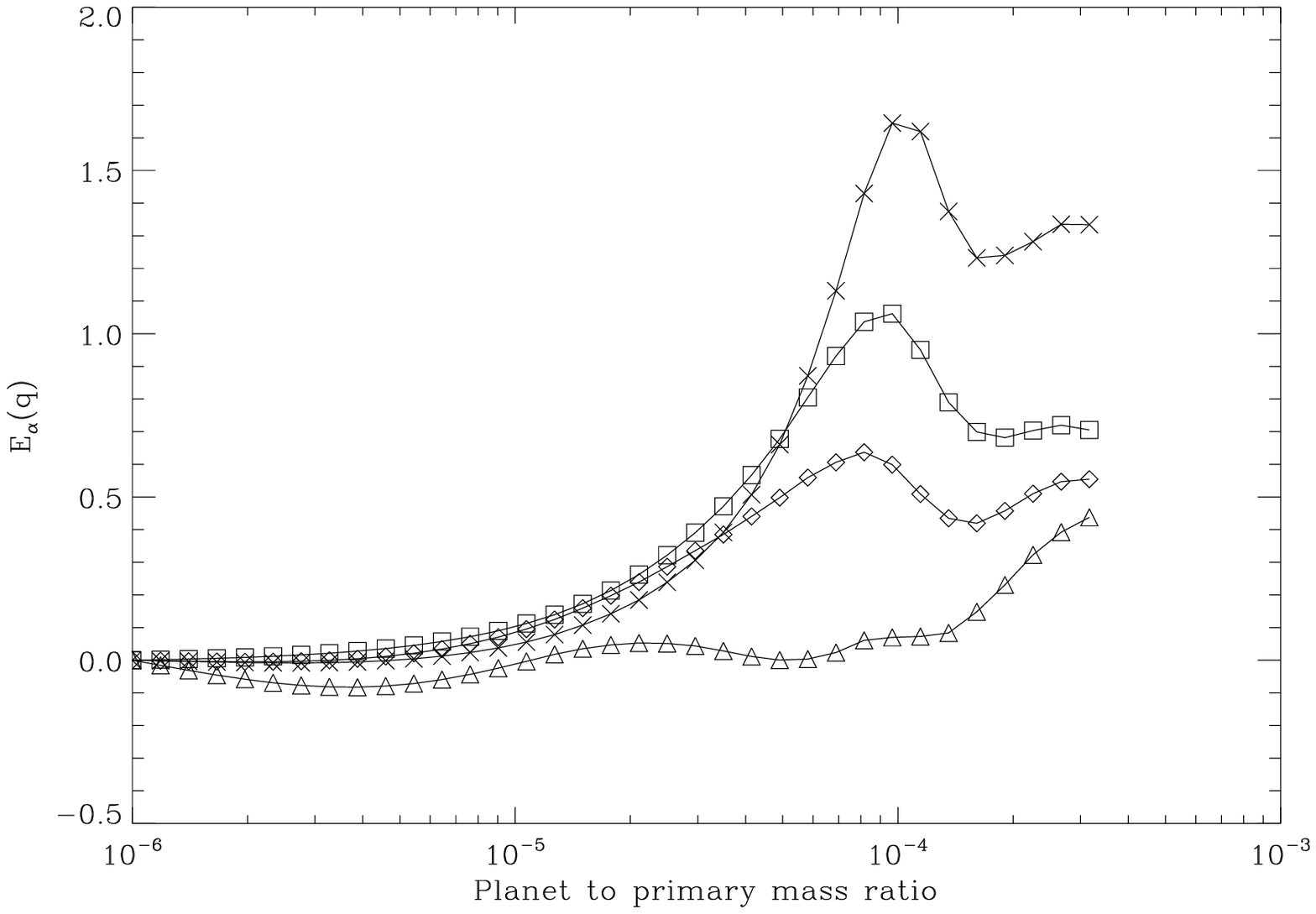}
\caption{\label{fig:fargoviscdep}Left:  Specific torque acting  on the
  planet  as  a  function  of  the  planet  mass  for  different  disk
  viscosities:   $\nu=5\cdot   10^{-6}$   (triangles),   $\nu=10^{-5}$
  (reference    calculations,    diamonds),    $\nu=2\cdot    10^{-5}$
  (squares). The curve with crosses  shows the torque of the reference
  calculations averaged between $t=6$ and $t=7$~orbits, i.e. the early
  value of the torque, before  it possibly saturates. Right: The value
  of $E_\alpha$ (given by Eq.~\ref{eqn:ealpha}) for these calculations
  (same symbols).}
\end{figure}

We  have  taken  twice  the  viscosity  reference  value  ($\nu=2\cdot
10^{-5}$), and  half the  reference value ($\nu=5\cdot  10^{-6}$). The
results are presented in Fig.~\ref{fig:fargoviscdep}.
The trend  observed on this  figure is compatible with  the saturation
properties of  the corotation torque.  The largest  offset is observed
for the  early torque  value, i.e. the  unsaturated one, while  as the
viscosity decreases  the departure  from linearity decreases  as well.
Quantitatively,  the behavior  observed is  also in  agreement  with a
corotation torque saturation.  The latter depends on the  ratio of the
libration timescale in the  horseshoe region and the viscous timescale
across it \citep{wlpi92,  m01, m02}. We can for  instance evaluate how
saturated  the  corotation  torque  should  be  for  $q=10^{-4}$.  The
horseshoe zone  half width $x_s$ for  such planet mass in  a disk with
$h=H/r=0.05$ can be  estimated by equating the linear  estimate of the
coorbital  corotation  torque \citep{tanaka}  and  the horseshoe  drag
\citep{wlpi91,wlpi92,m01}. One is led, in a two-dimensional disk, to:
\begin{equation}
\label{eqn:xs}
x_s=1.16a\sqrt\frac qh.
\end{equation}
This  yields  here  $x_s=0.052$.   The  ratio ${\cal  R}$  defined  by
\citet{m01} is therefore ${\cal R}=0.07$ for the reference run, ${\cal
  R}=0.14$ for the larger viscosity  run, and ${\cal R}=0.035$ for the
lower  viscosity  run.   To  within  a numerical  factor,  ${\cal  R}$
represents  the  ratio  of  the  libration timescale  to  the  viscous
timescale across the horseshoe region, and therefore indicates whether
the corotation  torque should saturate  (at low ${\cal R}$)  or remain
unsaturated (at  higher ${\cal R}$).  From Fig.~2  of \citet{m02}, one
can infer that the coorbital corotation torque should be about $40$~\%
of  its  unsaturated  value  for the  smaller  viscosity  calculation,
$60$~\%  for the  reference calculation,  and $80$~\%  for  the larger
viscosity    calculation.    The    scatter   of    the    curves   of
Fig.~\ref{fig:fargoviscdep}b   is   roughly   compatible  with   these
expectations. We  note in  passing that (i)  this estimate is  only an
order of magnitude estimate, since we inferred the value of $x_s$ from
linear  calculations, whereas we  suspect the  offset to  be due  to a
corotation torque value that differs from the linear estimate and (ii)
it  is by  chance  that  the reference  calculation,  which takes  the
parameters of \citet{dkh03}  and \citet{bate03}, corresponds precisely
to  a  corotation torque  that  is  half  saturated, so  that  varying
slightly  the viscosity  with respect  to the  reference one  yields a
strong variation  of the offset  amplitude.  We finally note  that the
saturation of the corotation torque  depends on the planet mass, for a
fixed viscosity.  The  smaller the planet mass, the  less saturated is
the    corotation    torque.    We    observe    this   behavior    in
Fig.~\ref{fig:fargoviscdep}b.  Quite  surprisingly however, the torque
is  found to  depend  (weakly) on  the  viscosity at  very small  $q$,
whereas one would expect the corotation torque to be unsaturated.  The
evolution of  the surface density profile  is too weak  to account for
this  observation.  We  have not  investigated further  this behavior,
which  we believe to  be of  minor importance  for the  work presented
here. Nevertheless,  we suggest  that it  is linked to  a drop  of the
coorbital corotation  observed by  \citet{m02}, when the  viscosity is
larger than the so-called  cut-off viscosity, which corresponds to the
viscosity for which  the time needed by a fluid  element to drift from
the separatrix  to the corotation is  also half the  libration time of
this  fluid  element.   This  limit  viscosity  $\nu_l$  is  given  by
$\nu_l\sim       x_s^2\Omega_p/4\pi$      \citep{m01,m02}.       Using
Eq.~(\ref{eqn:xs}),    this    translates    into    $\nu_l\sim    0.1
a^2\Omega_pq/h$.  We should observe  a drop  of the  corotation torque
(and  therefore a  dependence  of  the torque  on  the viscosity)  for
$\nu\ga\nu_l$,  {\em  i.e.}   for $q\la  q_l\sim10h\nu/(a^2\Omega_p)$.
For the reference calculation, we have $q_l\sim 5\cdot 10^{-6}$, while
we get  twice and half this  value for the higher  and lower viscosity
runs,  respectively. The  curves  of Fig.~\ref{fig:fargoviscdep}a  are
roughly  compatible with  these expectations,  although  around $q\sim
10^{-5}$ it is difficult to  disentangle this effect from the onset of
the departure from linearity of the torque.

\subsubsection{3D results}
\begin{figure}
\plotone{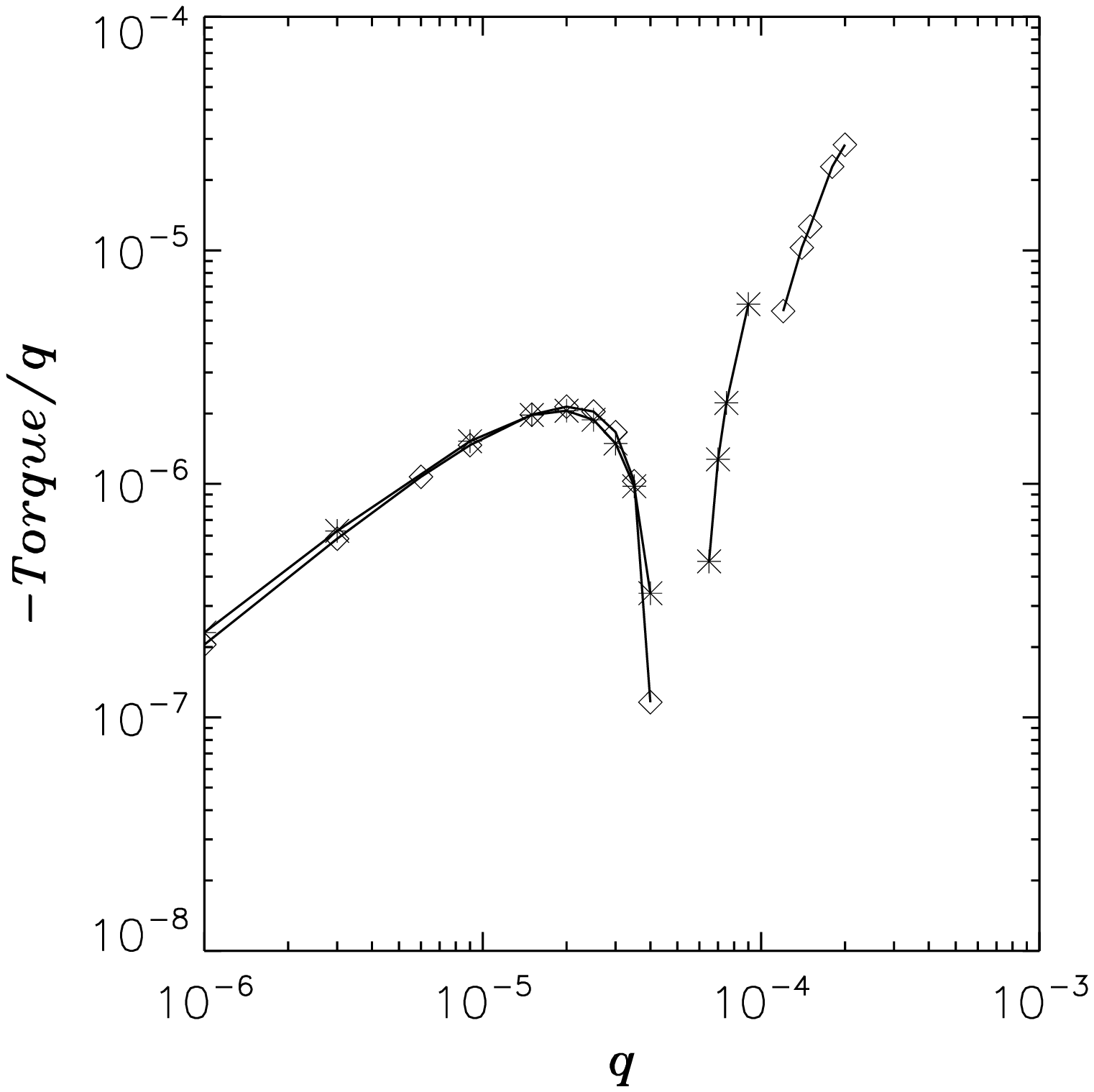}
\caption{\label{fig:3dviscdep} Negative of  the specific torque acting
  on  the  planet,  as  a  function  of  its  mass,  for  models  with
  $\alpha=0$. Asterisks  indicate torques, from  long-run simulations,
  for  which  corotation torques  are  saturated.  Diamonds  represent
  torques  at  early  times  (between  $7$  and  $10$  orbits),  hence
  corotation torques are unsaturated.}
\end{figure}
As explained above, torques evaluated at early times contain coorbital
corotation torques that  are unsaturated and thus their  effect is the
strongest.  At later  evolutionary  times, the  effects of  corotation
torques  may tend  to weaken.   Figure~\ref{fig:3dviscdep} illustrates
the behavior of saturation on the total specific torque, as a function
of the  planet mass,  obtained from calculations  with a  flat initial
surface  density   ($\alpha=0$).   The  asterisks   represent  torques
measured around  $100$ orbits,  when corotation torques  are partially
saturated whereas  diamonds refer to torques measured  between $7$ and
$10$  orbits,  before  saturation   occurs.   The  offset  reduces  as
corotation torques saturate.  The planet  mass for which the offset is
maximum shifts towards larger values  and the range of masses in which
the total  torque is positive  shrinks (see Fig.~\ref{fig:3dviscdep}).
However, a finite  mass interval persists in which  the departure from
the linear regime can still be very large.

\subsection{Dependence on the disk thickness}
\label{sec:depthick}
The two previous sections strongly suggest that the offset is indeed a
physical effect, independent  on the code used, and  that it is linked
to an  excess of the coorbital  corotation torque with  respect to its
linearly  estimated  value. This  therefore  implies  that the  offset
corresponds to the onset of  non-linear effects in the flow.  The flow
non-linearity   depends   on   the  parameter   ${\cal   M}=q^{1/3}/h$
\citep{kpap96}.  The  onset  of  this  behavior  should  therefore  be
observed for a  planet to primary mass ratio  $q\propto h^3$.  We have
undertaken  additional series  of calculations  in which  we  take the
reference parameters of Table~\ref{tab:reference}, except that we vary
the disk aspect ratio.

\subsubsection{2D results}
\begin{figure}
\plotone{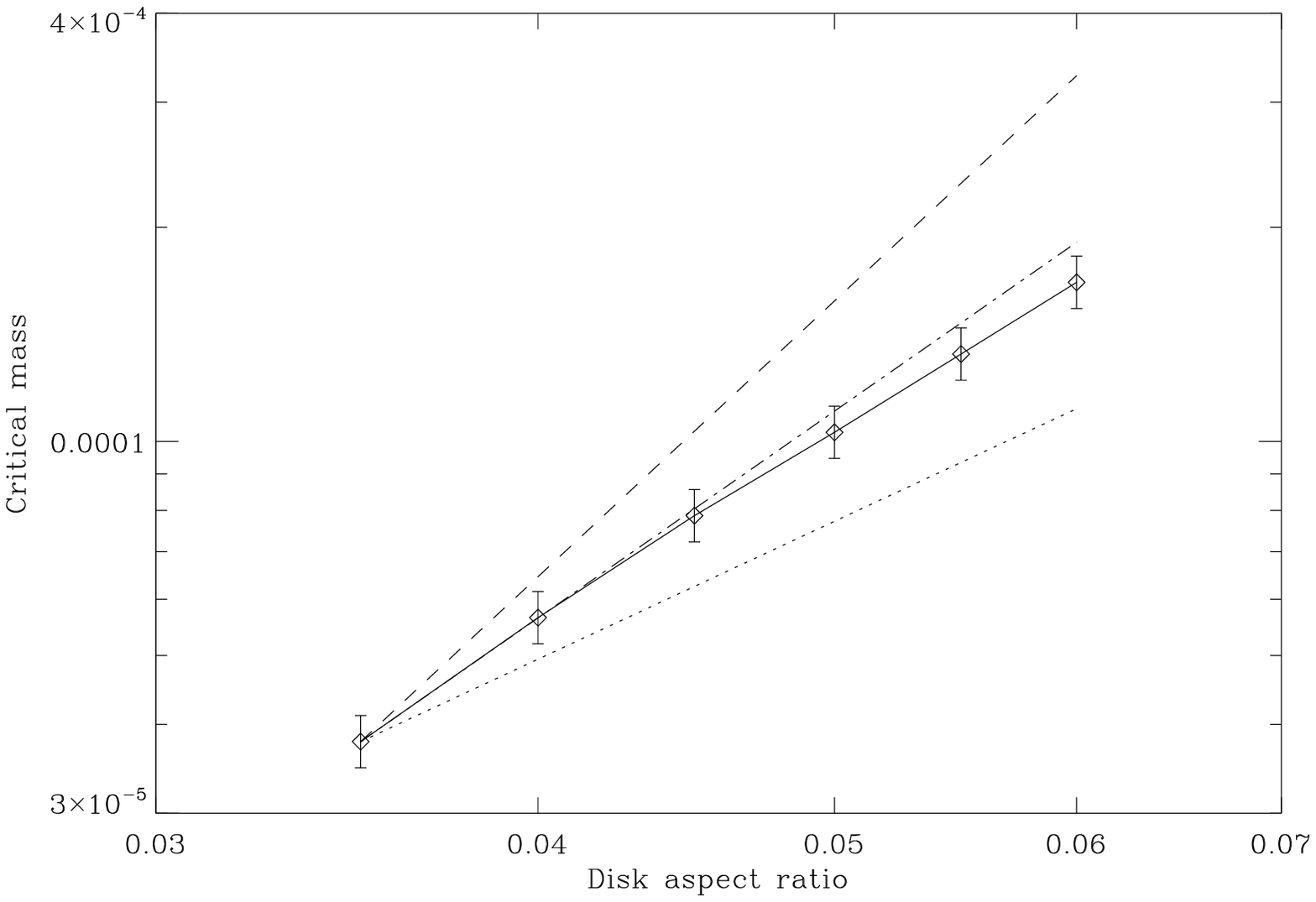}
\caption{\label{fig:fargodeph}Critical  mass $q_c$ for  maximal offset
  as a function  of the disk thickness, for the  2D runs.  The dotted,
  dashed  and  dot-dashed lines  show  respectively the  relationships
  $q_c\propto h^2$,  $q_c\propto h^4$  and $q_c\propto h^3$  that pass
  through  the  leftmost  data  point.  The error  bars  indicate  the
  sampling of data points around the critical mass.}
\end{figure}

We  ran series  of calculations  with $h=0.035$,  $h=0.04$, $h=0.045$,
$h=0.055$ and $h=0.06$, in  addition to the reference calculation with
$h=0.05$.   For  each series,  we  estimate  the  mass for  which  the
departure to  linearity given  by Eq.~\ref{eqn:ealpha} is  maximal. We
refer to this mass as the critical mass, and we denote $q_c$ its ratio
to  the  primary mass.   This  mass  is  determined from  a  parabolic
interpolation of  the data point  which has the largest  departure and
its two neighbors. Since the disk viscosity is kept constant and equal
to  its reference  value  in  all these  calculations,  and since  the
critical  mass varies  between  two sets  of  calculations, we  expect
different saturation levels of  the coorbital corotation torque at the
critical mass, on the long  term.  This could mangle our analysis, and
it is therefore  important to take the unsaturated  torque value. This
is why the $E_{\alpha}(q)$ values  in the analysis of this section are
evaluated  using   the  torque   value  averaged  between   $t=3$  and
$t=5$~orbits. The results are presented in Fig.~\ref{fig:fargodeph}.
We see on this figure that there is an excellent agreement between the
results  of the  calculations  and the  expectation $q_c\propto  h^3$.
This is a  strong point in favor of our  hypothesis that this behavior
is due to the onset of non-linear effects.

\subsubsection{3D results}
\begin{figure}
\plotone{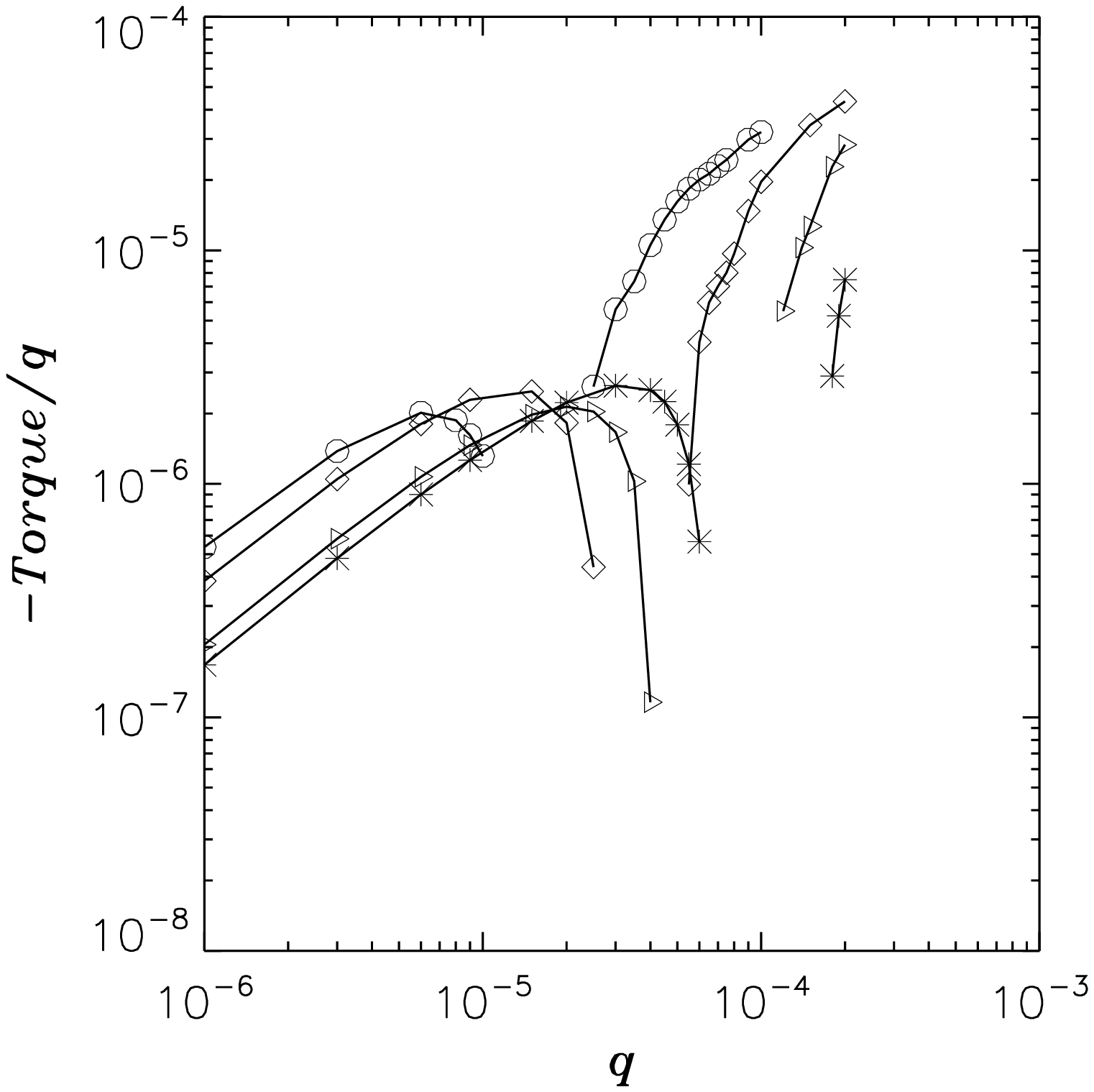}
\caption{\label{fig:3ddepqh} Negative of the specific torque acting on
  the planet  as a function of  the planet to primary  mass ratio, for
  different  values of  $h$: $0.06$  (asterisks),  $0.05$ (triangles),
  $0.04$ (diamonds),  and $0.03$  (circles).  Torques are  measured at
  early times (between $7$ and $10$ orbits) so that corotation torques
  are  unsaturated.   Gaps  in  the  curves  identify  the  ranges  of
  planetary masses for which the total torque is positive.}
\end{figure}
In order  to examine the dependence  of the offset on  the disk aspect
ratio, we set up 3D models with an initial surface density slope equal
to $\alpha=0$ and a relative disk thickness $h$ ranging from $0.03$ to
$0.06$.  For  each value  of $h$,  a series was  built by  varying the
planet to  star mass ratio,  $q$, from $10^{-6}$  to $2\times10^{-4}$.
The specific  torque as  a function of  the planet mass,  for selected
disk aspect  ratios, is shown in  Figure~\ref{fig:3ddepqh}. In thinner
(i.e., colder)  disks, the offset of corotation  torques moves towards
smaller planetary cores. When $h=0.03$,  the effects of the offset are
dominant  between  $q\approx10^{-5}$  (or about  $3\;M_{\oplus}$)  and
$q\approx2\times10^{-5}$ (or about $6\;M_{\oplus}$), regardless of the
saturation level of corotation torques.
\begin{figure}
\plotone{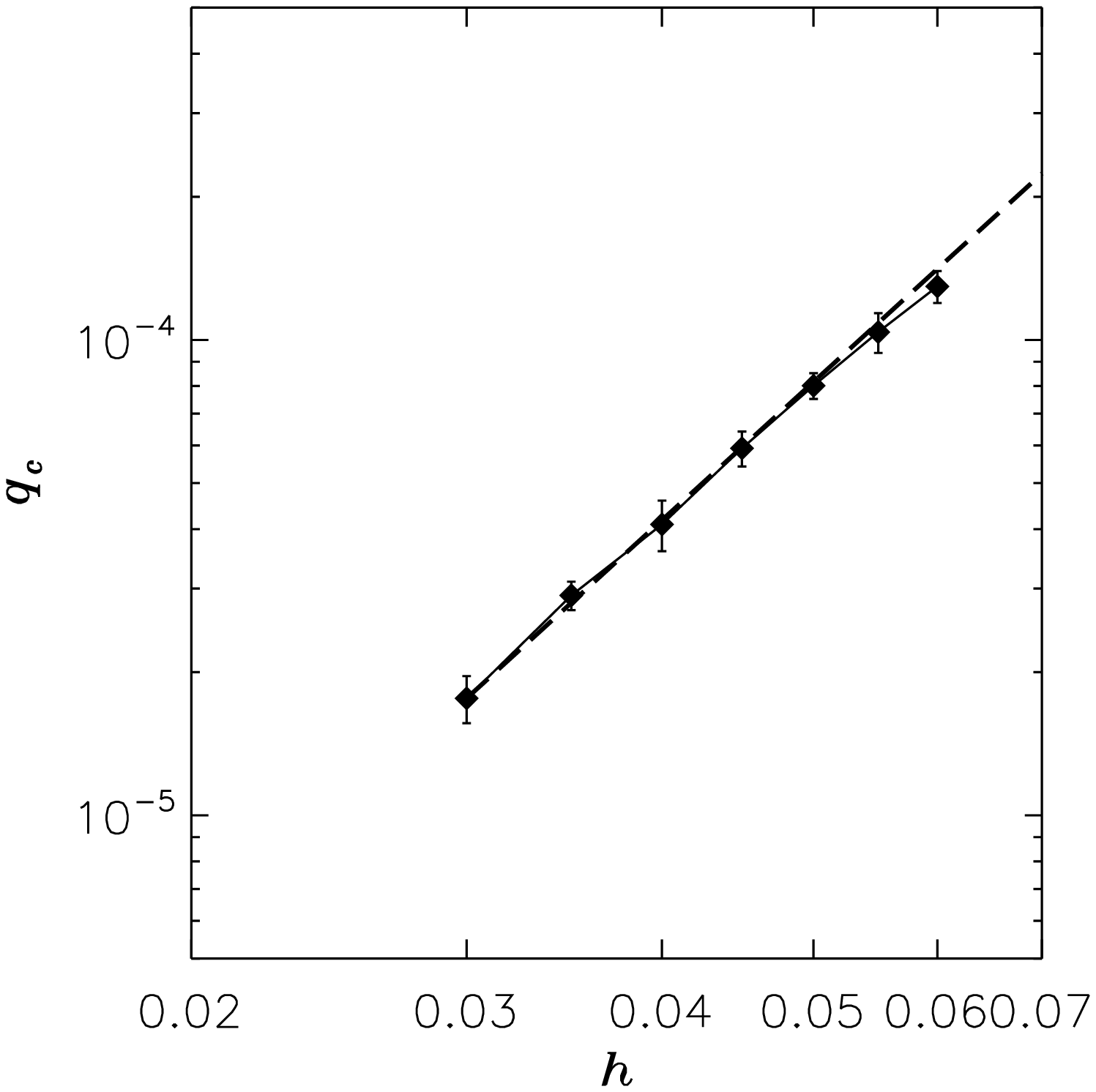}
\caption{\label{fig:3ddeph}Critical  mass  versus  the  relative  disk
  thickness obtained from 3D calculations. At $q=q_{c}$, the offset is
  the largest. The error bars indicate the sampling of the data points
  around  the   critical  mass.    The  dashed  line   identifies  the
  relationship $q_c\propto  h^3$ passing  through the data  point with
  minimum $h$.}
\end{figure}
From each  series, the  critical mass ratio  $q_{c}$ was  estimated by
means of a  parabolic interpolation, as done for  the 2D calculations.
For this  analysis we  used total torques  averaged between  $t=7$ and
$t=10$ orbits,  i.e. before  the corotation torque  possibly saturate,
for the reasons clarified in  the previous section.  The dependence of
the  critical mass  ratio  on  the disk  thickness  is illustrated  in
Figure~\ref{fig:3ddeph},       along        with       the       curve
$q_{c}/q_{c}(h=0.03)=(h/0.03)^3$   (dashed  line).   The   error  bars
indicate the sampling of the  data points around the critical mass and
thus  represent  the  largest  possible  error  on  the  estimates  of
$q_{c}$. It is evident  that 3D numerical results accurately reproduce
the  $h^3$-scaling expected to  arise from  non-linear effects  in the
corotation region.

\section{Streamline analysis}
\label{sec:stream}
\begin{figure}
\plotone{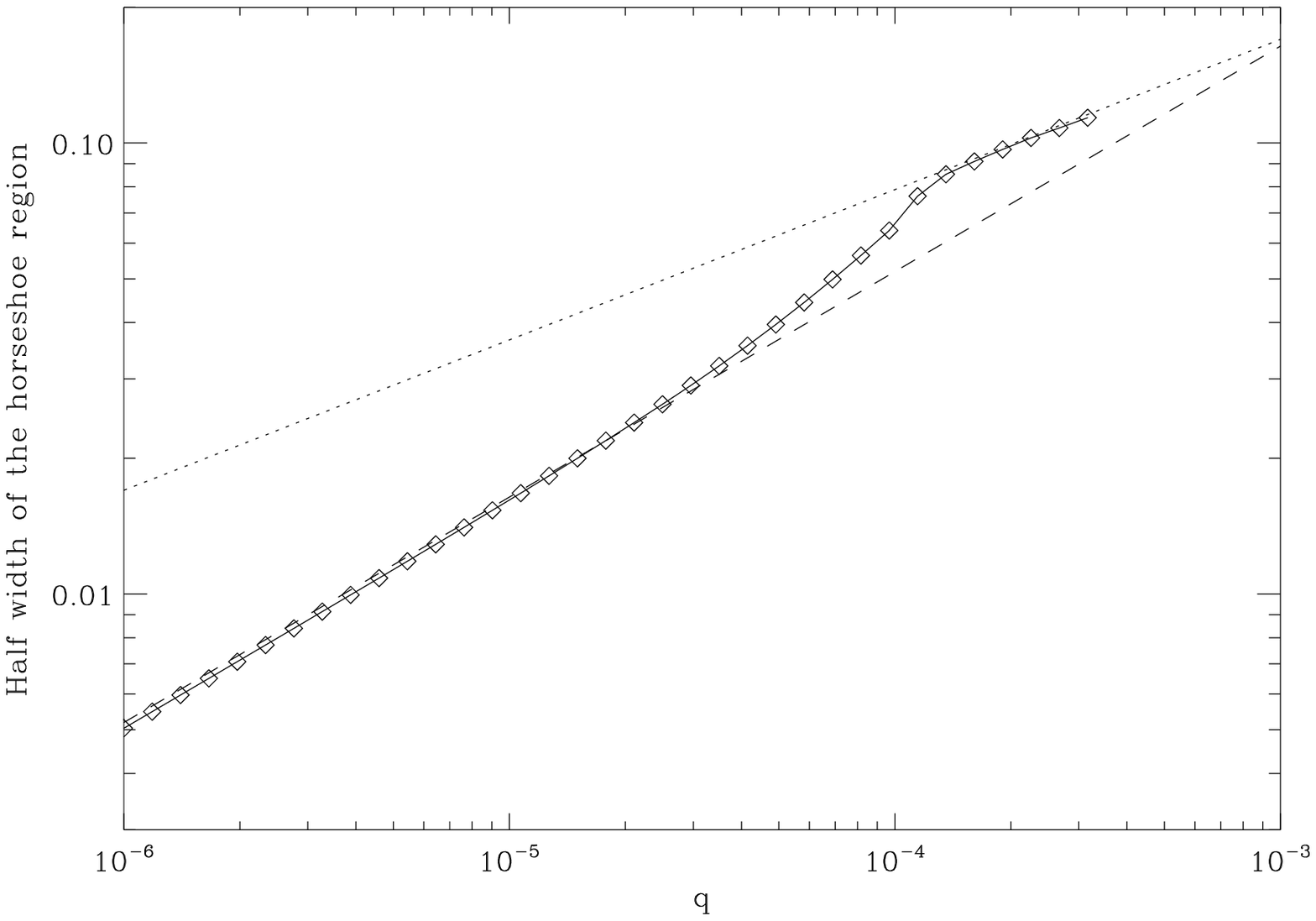}
\caption{\label{fig:hswidth}Horseshoe zone half width as a function of
  the planet mass for the  $35$ planets of the reference calculations.
  The dashed  line represents the  horseshoe zone half  width expected
  from Eq.~(\ref{eqn:xs}).  It scales with $q^{1/2}$.  The dotted line
  shows the relationship $x_s\propto q^{1/3} $ that passes through the
  large mass data points.}
\end{figure}

The calculations  shown at the previous section  strongly suggest that
the offset is a physical effect, and that non-linear effects boost the
corotation  torque  value  with  respect  to  its  linearly  estimated
value. There is a link between the coorbital corotation torque and the
so-called  horseshoe drag~\citep{wlpi91,wlpi92,m01,m02}, which  is the
torque arising  from all the  fluid elements of the  horseshoe region.
Although  the  corotation torque  and  the  horseshoe  drag have  same
dependency  on  the  disk  and  planet parameters,  and  although  the
horseshoe drag may result in a very effective concept for some aspects
of  planetary migration  related to  coorbital  material \citep{mp03},
there  is no reason  why these  two quantities  should be  exactly the
same. In particular, in the  low mass regime, the horseshoe region can
be  arbitrarily radially  narrow, while  the corotation  torque always
arises, in  the linear limit, from  a region of width  $\sim H$, which
corresponds  to the length-scale  over which  the disturbances  in the
corotation vicinity  are damped.   Nevertheless, it is  instructive to
investigate whether  the behavior  found is linked  to a boost  of the
horseshoe region width w.r.t.  its linearly estimated width. We recall
the horseshoe drag expression \citep{wlpi91,wlpi92,m01}:

\begin{equation}
\label{eqn:hsdrag}
\Gamma_{HS}=\frac 34x_s^4\Omega_p^2\Sigma\cdot
\frac{d\log(\Sigma/B)}{d\log r},
\end{equation}

where $x_s$ is  the half width of the  horseshoe region, $\Omega_p$ is
the planet orbital frequency and  $\Sigma$ is the disk surface density
at the orbit.  Since, in the  linear limit, the torque scales with the
square  of  the planet  mass,  we  expect  the dependency  $x_s\propto
q^{1/2}$  (see also  \citet{wlpi92}). On  the large  mass side  we may
expect, that the horseshoe region has a behavior similar to the one of
the restricted three body problem  (RTBP) and that we have the scaling
$x_s\propto q^{1/3}$.   We performed an  automatic streamline analysis
on the flow of the 2D reference runs%
\footnote{The  runs on  which  the streamline  analysis was  performed
  differ slightly from  the reference runs of section~\ref{sec:setup}:
  (i)  the  resolution  was  increased,  with  $N_{\rm  rad}=386$  and
  $N_\phi=1728$, and  the radial  interval was narrowed,  from $R_{\rm
    min}=0.6$ to  $R_{\rm max}=2.0$; (ii) the sound  speed, instead of
  the  aspect  ratio,  was   taken  uniform,  so  that  $H(r=1)=0.05$.
  Everything else  corresponds to the  reference runs.}, in  the frame
corotating  with the planet,  after $t=10$~orbits  (an early  stage in
order to avoid, on the large mass side, a radial redistribution of the
disk material that alters the streamlines and hence the horseshoe zone
width,  but  still  sufficiently  evolved  so that  the  flow  can  be
considered steady with a  good approximation in the corotating frame),
in  order  to find  the  separatrices of  the  horseshoe  region by  a
bisection method.  We show in Fig.~\ref{fig:hswidth} the half width of
the horseshoe region as a function  of the planet mass. We see on this
figure that:
\begin{itemize}
\item the horseshoe  zone width indeed scales as  $q^{1/2}$ as long as
  the planet  mass remains sufficiently  small, since the  data points
  and the dashed line have same slope for $q<3\cdot 10^{-5}$;
\item there  is a correct  agreement between the  coorbital corotation
  torque and the horseshoe drag,  since the data points and the dashed
  curve,  obtained  from   Eq.~(\ref{eqn:xs})  by  assuming  a  strict
  equality  between horseshoe  drag and  linearly  estimated coorbital
  corotation torque, nearly coincide on this mass range.
\item We also  see how the horseshoe zone  width scales with $q^{1/3}$
  on  the large mass  side, as  expected. The  width displayed  on the
  dotted  line  however  differs  from  the  horseshoe  width  of  the
  RTBP. The latter is  $x_s=\sqrt{12}a(q/3)^{1/3}$, while we find that
  the    data   points    are   correctly    fitted    by   $x_s\simeq
  2.45a(q/3)^{1/3}$,  i.e. the  horseshoe  width is  $\sim 1.4$  times
  narrower than in the RTBP.
\item In  between the  linear range and  the $q^{1/3}$  scaling range,
  that is for $3\cdot  10^{-5}<q<1.5\cdot 10^{-4}$, the horseshoe zone
  width falls between the two  regimes, which makes it larger than its
  linearly estimated value for any $q>3\cdot10^{-5}$. This corresponds
  precisely  to  the mass  for  which  migration  becomes slower  than
  linearly estimated.
\end{itemize}
In order to finally assess whether the torque offset can indeed be due
to the  excess of the horseshoe  zone width, we  can directly estimate
the excess of horseshoe drag (w.r.t. the linearly extrapolated value):

\begin{equation}
\Delta\Gamma_{HS}(q)=\Gamma_{HS}(q)-\Gamma_{HS}(q_{min})\left(\frac{q}{q_{min}}
\right)^2,
\end{equation}
and compare it to the total torque excess:
\begin{equation}
\Delta T(q)=T(q)-T(q_{min})\left(\frac{q}{q_{min}}
\right)^2=\frac{|T(q_{min})|}{q_{min}^2}E_\alpha(q).
\end{equation}
\begin{figure}
\plotone{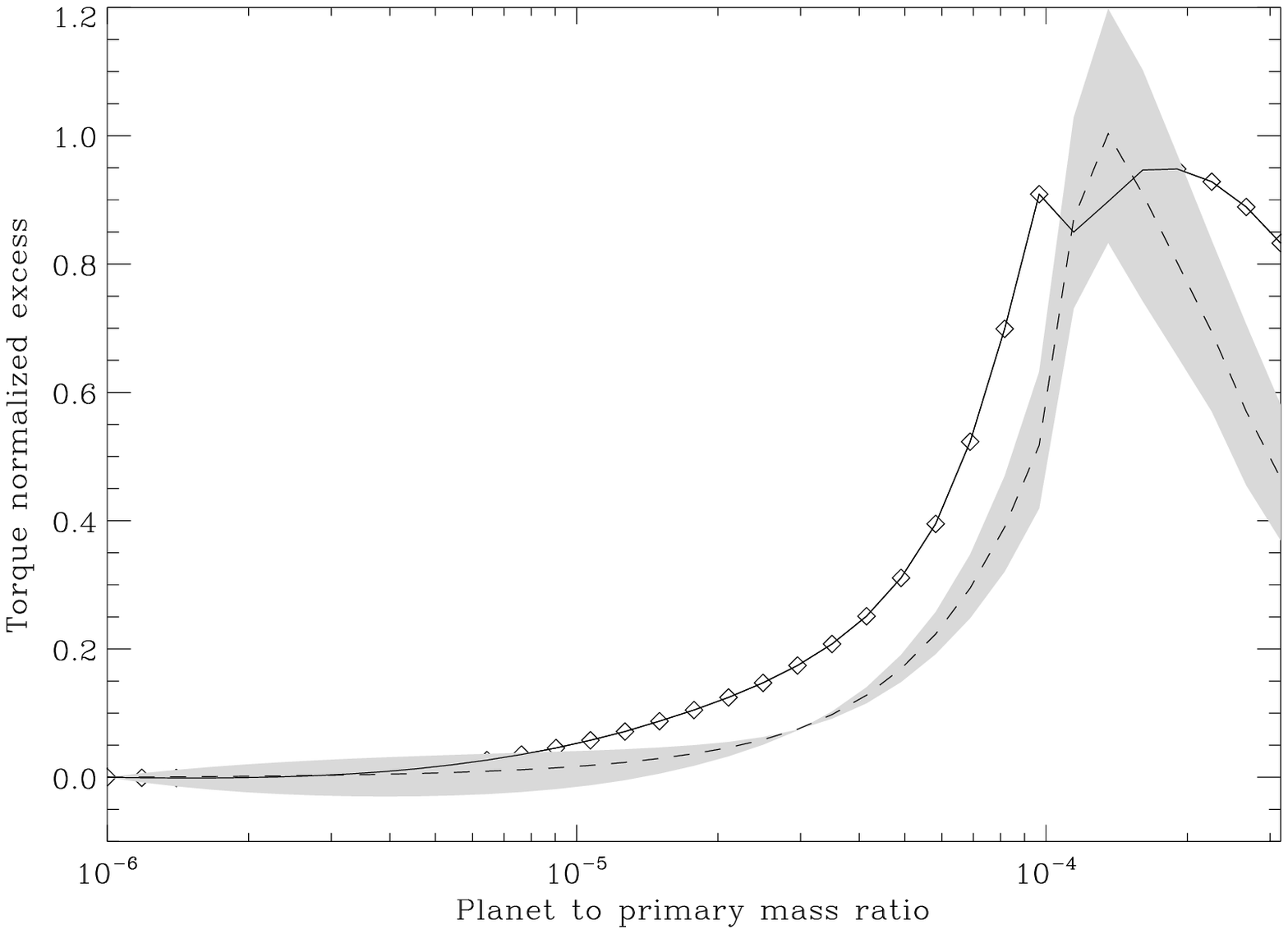}
\caption{\label{fig:tqvshs}Horseshoe  drag  (dashed  line)  and  total
  torque  (solid line)  normalized excesses  as a  function  of planet
  mass. The shaded  area shows the uncertainty on  the horseshoe drag,
  arising from  the uncertainty on  the horseshoe zone width.   If one
  calls  $x_s^-$ (resp.   $x_s^+$) the  distance of  the  inner (resp.
  outer) separatrix  to the corotation,  then the upper  (resp. lower)
  limit  of the  shaded  zone is  given  by using  $\max(x_s^-,x_s^+)$
  (resp.  $\min(x_s^-,x_s^+)$)  in  Eq.~(\ref{eqn:hsdrag}), while  the
  dashed line uses $(1/2)(x_s^-+x_s^+)$.}
\end{figure}
The results are displayed in Fig.~\ref{fig:tqvshs}, in which we divide
the torque values by $q^2$. We  see that the horseshoe drag excess and
the  total torque excess  exhibit the  same behavior  and have  a very
similar value in the  mass range $10^{-4}<q<2\cdot10^{-4}$, which is a
quantitative confirmation that the torque excess of the offset maximum
is  attributable to  the horseshoe  zone width  excess.  We  note that
although the  two curves display  a similar behavior  for $q<10^{-4}$,
they do  not coincide on  this mass range,  and that the  total torque
excess is systematically larger than the horseshoe drag excess.  It is
precisely  for  this   mass  range  ($q<10^{-4}\sim  h^3/1.16^2$,  see
Eq.~\ref{eqn:xs}) that  the horseshoe zone width is  narrower than the
disk thickness, so that not all the coorbital corotation torque arises
from the horseshoe region.

\section{Flow transition}
\label{sec:bernou}
\begin{figure}
\plotone{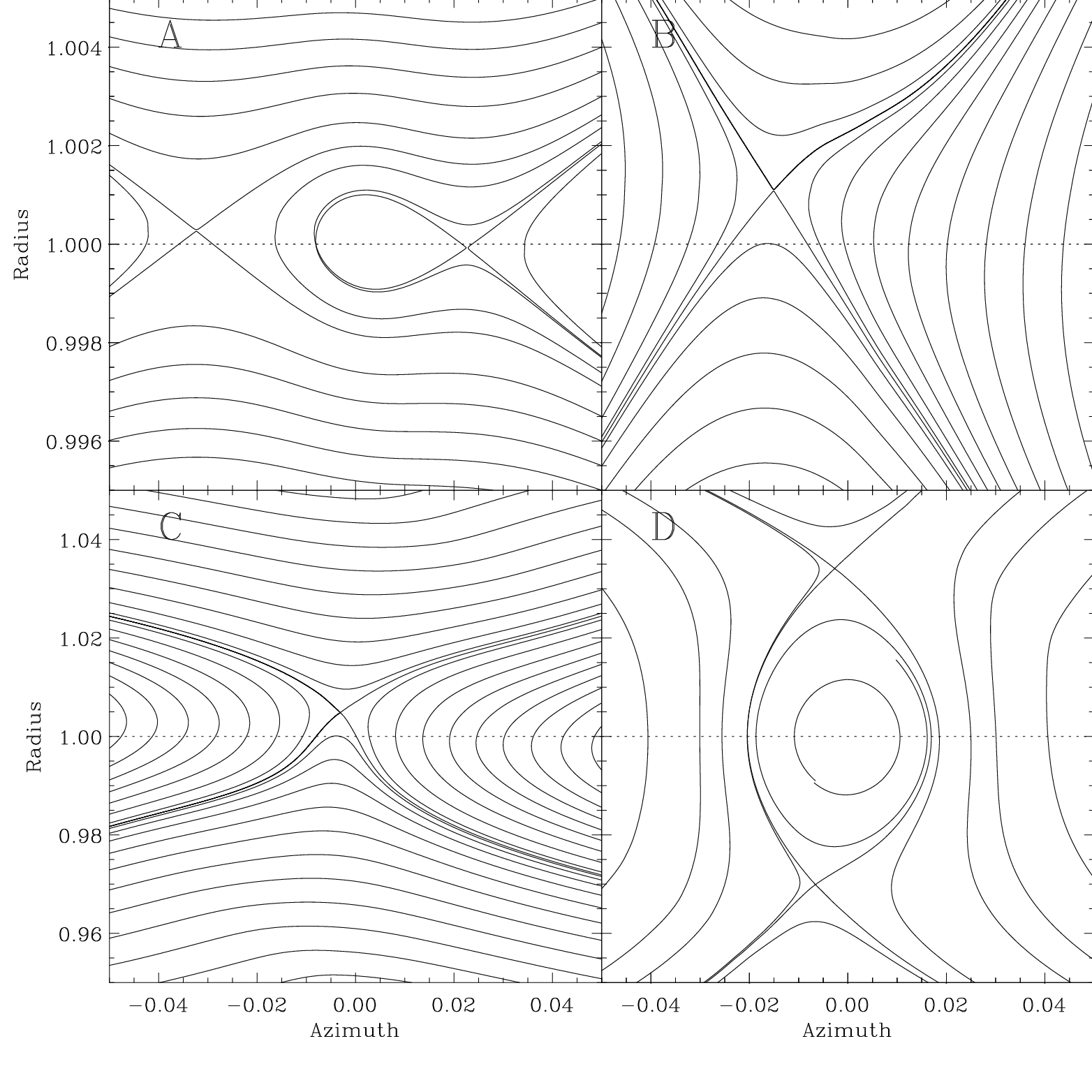}
\caption{\label{fig:flowtopo}  Streamline   appearance  for  the  four
  planet masses quoted in text,  at $t=10$~orbits. The radial range is
  the same for cases  A and B, and for cases C and  D. It is ten times
  larger  for the latter  than from  the former.  The aspect  ratio is
  $1:1$ for the cases C and D.}
\end{figure}

The  previous  section  shows that  the  torque  offset  is due  to  a
transition of the corotational flow,  which has a horseshoe zone width
$\propto q^{1/2}$ in the linear  regime whereas it scales as $q^{1/3}$
in   the  large  mass   regime.   Fig.~\ref{fig:flowtopo}   shows  the
streamline topology for different masses (A: $q=5.44\cdot 10^{-6}$; B:
$q=2.96\cdot  10^{-5}$;  C:  $q=8.16\cdot  10^{-5}$;  D:  $q=2.67\cdot
10^{-4}$).  The linear case (A) shows two stagnation points%
\footnote{We  restrict  ourselves to  the  case  of hyperbolic  points
  (X-type),  as  these  lie  on  the  separatrices  of  the  libration
  region. The  flow also features elliptic  stagnation points (O-type)
  such  as the  ones that  can be  found inside  the region  of closed
  streamlines in  case (A)  or (D). Since  those are not  connected to
  separatrices,  they are  not  relevant to  the present  discussion.}
located almost at  corotation, and offset in azimuth  from the planet.
These two stagnation points are  not symmetric w.r.t.  the planet, and
are not  located on  the same streamline.   As long  as we are  in the
linear regime, they remain essentially  at the same location. Then, as
the  planet mass increases,  both stagnation  points move  towards the
planet.  The central libration region defined by the separatrix of the
right stagnation point  shrinks until it disappears, in  which case we
only have  one stagnation  point (case B).   As the planet  mass still
increases, this unique stagnation  point moves towards smaller azimuth
while it  recedes radially  from the orbit  (case C), then  for larger
masses one  gets two stagnation points practically  on the star-planet
axis,  which yields  a picture  very similar  to the  RTBP,  where the
stagnation  points are reminiscent  of the  Lagrange points  L$_1$ and
L$_2$, and a prograde  circumplanetary disk appears within the ``Roche
lobe'' (case D). This corresponds to the regime in which the horseshoe
zone width scales with $q^{1/3}$.

\begin{figure}
\plotone{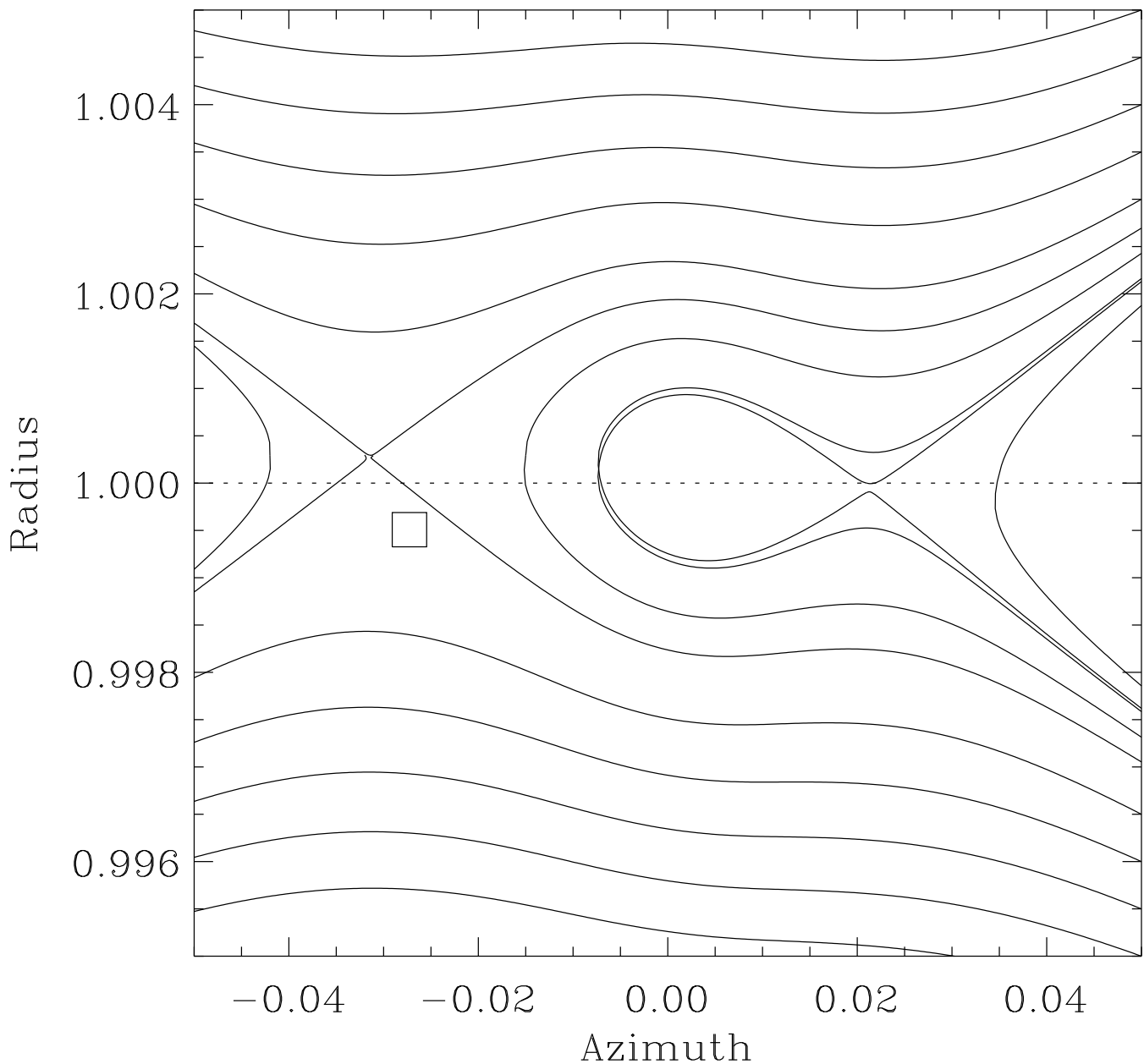}
\caption{\label{fig:flowtopohr}  Streamline appearance for  the planet
  mass A quoted in text, at $t=10$~orbit, with a radial resolution ten
  times  higher  than   in  Fig.~\ref{fig:flowtopo}.  The  streamlines
  appearance  and the  position  of the  stagnation  points is  almost
  indistinguishable from the lower resolution case. The square shows a
  zone from the high resolution mesh.}
\end{figure}

One could  argue that  despite the larger  resolution adopted  for the
streamline   analysis,  the   radial   resolution  $\delta   r=(R_{\rm
  max}-R_{\rm min})/N_{\rm rad}=3.63\cdot 10^{-3}$ is still too coarse
to properly describe the corotational  flow of the small mass planets,
as it amounts  to a significant fraction of  the horseshoe zone width.
Fig.~\ref{fig:flowtopohr}  shows the  flow  for $q=5.44\cdot  10^{-6}$
(case  A) run  with  ten  times higher  a  radial resolution  ($N_{\rm
  rad}=3860$,  hence  $\delta  r=3.63\cdot 10^{-4}$).   The  excellent
agreement  between the  streamlines  obtained with  the two  different
radial resolutions confirms a  fact already noted by \citet{m02}, that
even  a low  or  mild  radial resolution  associated  with a  bilinear
interpolation of  the velocity fields allows to  capture correctly the
features of the corotation region.

\begin{figure}
\plotone{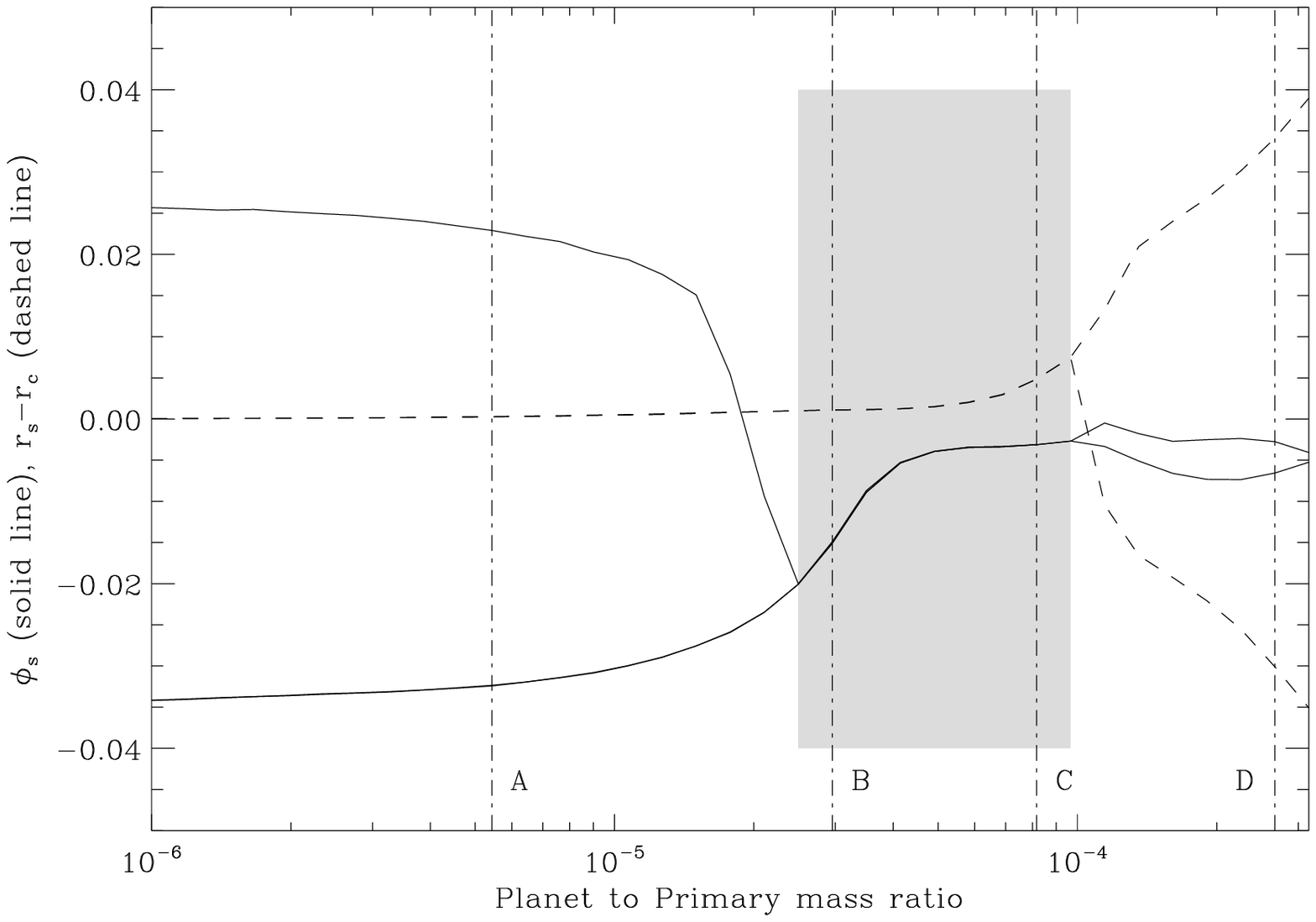}
\caption{\label{fig:stagpoints}  Azimuth  $\phi_s$  of the  stagnation
  point(s) (solid line) and  radial distance to corotation ($r_s-r_c$)
  of  the fixed point(s)  (dashed line)  as a  function of  the planet
  mass. The gray shaded zone  shows the mass interval over which there
  is  a unique stagnation  point. The  four vertical  dot-dashed lines
  show  the  masses  for  which  the  flow  topology  is  sketched  in
  Fig.~\ref{fig:flowtopo}.}
\end{figure}

These  flow properties  are illustrated  in Fig.~\ref{fig:stagpoints},
which shows  both the  azimuth and the  distance to corotation  of the
stagnation point(s).  We  see that for $q<2\cdot 10^{-5}$  we have two
stagnation points located at corotation and on each side of the planet
(i.e. one at  negative azimuth, and one at  positive azimuth).  Around
$q\sim  2\cdot 10^{-5}$, the  stagnation points  coalesce on  a narrow
mass interval.   Up to $q\sim  10^{-4}$, there is a  unique stagnation
point  located slightly  beyond corotation  and at  a  small, negative
azimuth.  Finally, at  $q\approx 10^{-4}$, another bifurcation occurs,
and one recovers  two stagnation points on either  side of corotation,
and almost aligned with the star ($|\phi_s|\ll|r_s-r_c|/a$).

For a given  finite potential softening length, there  is a mass limit
under  which  a 2D  flow  is linear  everywhere,  even  at the  planet
location.   A simple  estimate  of this  mass  limit can  be found  as
follows.  The  effective potential that  dictates the motion  of fluid
elements is $\tilde\Phi=\Phi+\eta$,  where $\Phi$ is the gravitational
potential and  $\eta$ is the  gas specific enthalpy. The  latter reads
$\eta=\eta_0+\eta'$, where $\eta_0$ is  the fluid specific enthalpy of
the  unperturbed flow, which  is a  uniform quantity  as the  disk has
initially a  uniform sound  speed and a  uniform surface  density, and
where  $\eta'=c_s^2\log(\Sigma/\Sigma_0)$ is  the perturbation  of the
specific   enthalpy  introduced   by  the   planet.    Similarly,  the
gravitational potential can  be written as $\Phi=\Phi_*+\Phi_p$, where
$\Phi_*$, the gravitational potential of the central star, corresponds
to  the  unperturbed  flow   and  where  $\Phi_p$,  the  gravitational
potential of  the planet, corresponds  to the perturbation.  Hence the
effective  potential can  be  decomposed as  $\tilde\Phi=\tilde\Phi_0+
\tilde\Phi'$, where  $\tilde\Phi_0=\Phi_*+\eta_0$ is its  value in the
unperturbed flow while $\tilde\Phi'=\Phi_p+\eta'$ is its perturbed
value.

\begin{figure}
\plotone{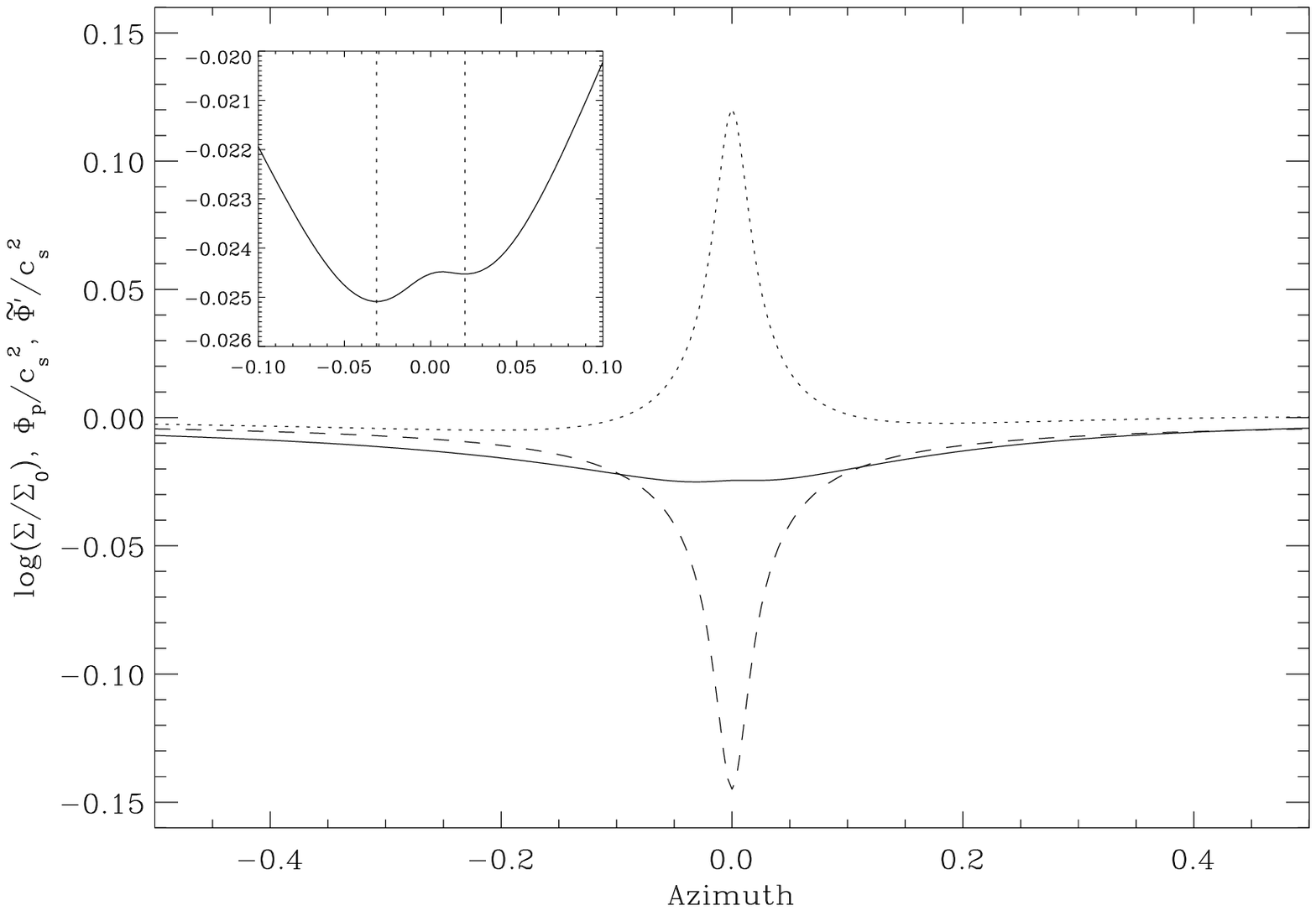}
\caption{\label{fig:enthalpy}This  graph shows $\log(\Sigma/\Sigma_0)$
  (dotted   line),  $\Phi_p/c_s^2$   (dashed  line)   and   their  sum
  ($\tilde\Phi'/c_s^2$, solid line) as  a function of azimuth at $r=1$
  for the  case A  (planet mass $q=5.44\cdot  10^{-6}$) with  the very
  high  radial resolution  ($N_{\rm rad}=3860$).   We note  in passing
  that  the close  up shows  two relative  extrema (shown  by vertical
  dotted  lines) which correspond  to the  position of  the stagnation
  points  shown  in  Fig.~\ref{fig:flowtopo}.   The  Bondi  radius  to
  softening length ratio for this planet is $r_B/\epsilon=0.145\ll 1$,
  which implies that  the flow is linear even  at the planet location.
  We     see     that      indeed     the     maximum     value     of
  $|\Sigma-\Sigma_0|/\Sigma_0\approx \log(\Sigma/\Sigma_0)$  is of the
  order of the above ratio.}
\end{figure}

\begin{figure}
\plotone{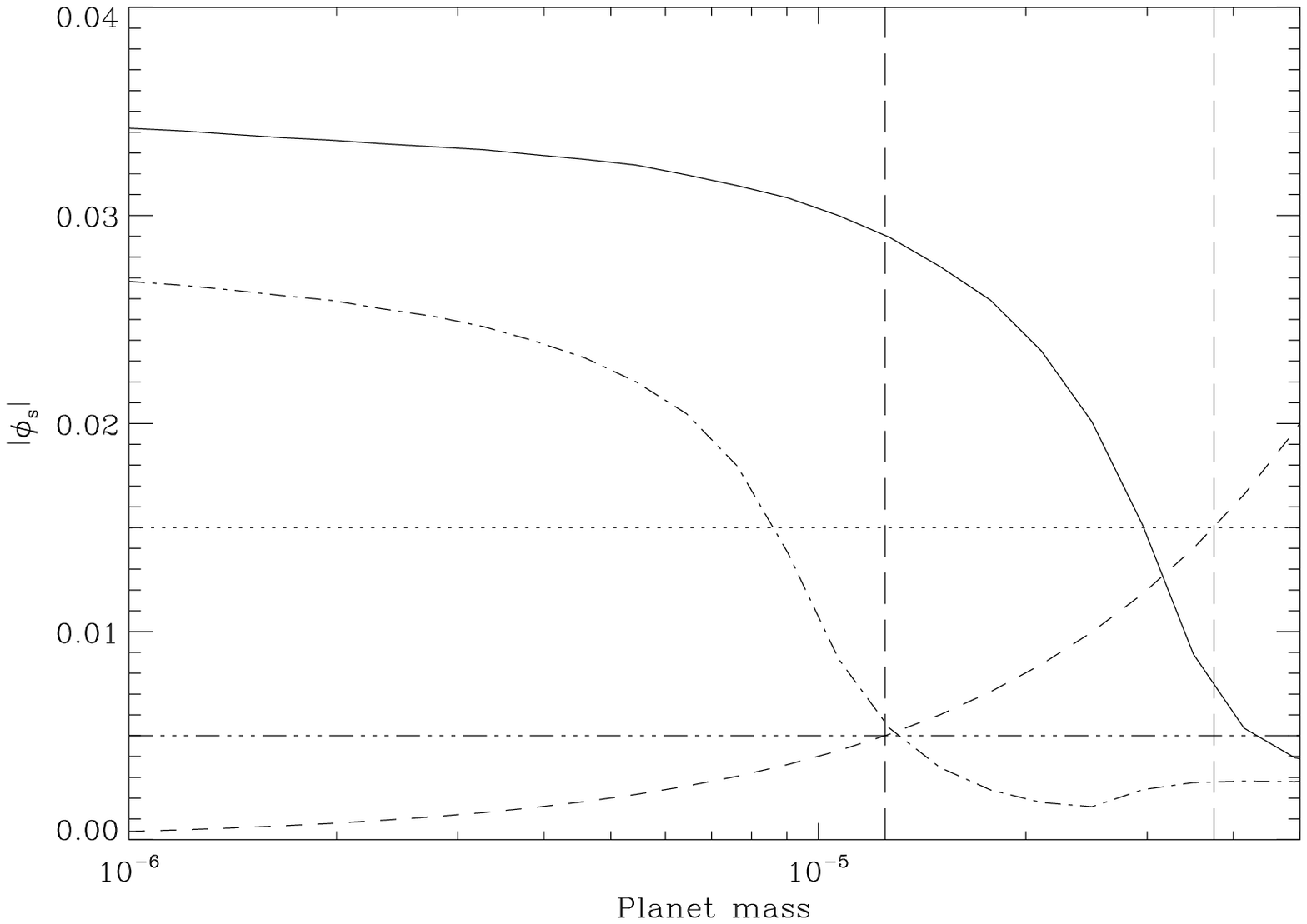}
\caption{\label{fig:diffbondi}Negative  of  the  azimuth of  the  left
  stagnation point as  a function of the planet  mass for the standard
  softening length case ($\epsilon=0.3H$,  solid line) and for a three
  times  smaller softening  length case  ($\epsilon=0.1H$, dash-dotted
  line). The dashed  line shows the Bondi radius as  a function of the
  planet mass. The horizontal dotted line shows the standard softening
  length ($\epsilon=0.015$) while the horizontal three-dot-dashed line
  shows the shorter softening length ($\epsilon=0.005$).  The vertical
  lines show  the mass for which  the planet's Bondi radius  is equal to
  the  potential softening  length  in  both cases.  We  see that  the
  stagnation point enters the Bondi sphere when $r_B\approx\epsilon$.}
\end{figure}

Figure~\ref{fig:enthalpy} shows  that the two  quantities $\Phi_p$ and
$\eta'$ are of the same order of magnitude and of opposite sign in the
planet vicinity, so that  the perturbed effective potential reduces to
a tiny fraction of the absolute value of either quantity.  A condition
for  the flow  linearity is  that $|\Sigma-\Sigma_0|/\Sigma_0  \ll 1$,
which therefore translates into $|\eta'|/c_s^2\ll 1$, or, at the planet
location, into:
\begin{equation}
r_B\ll \epsilon,
\end{equation}
where
\begin{equation}
r_B=\frac{GM_p}{c_s^2}
\end{equation}
is the planet's Bondi radius.  The flow linearity in the planet vicinity
in  a 2D  calculation  is therefore  controlled  by the  ratio of  the
potential softening length to the Bondi radius.
Fig.~\ref{fig:diffbondi} shows  the absolute  value of the  azimuth of
the  left  stagnation  point as  a  function  of  mass, for  the  runs
described below  as well as for a  similar set of runs  with a smaller
softening length ($\epsilon=0.1H=0.005$).  In  both cases, we see that
as  long  as  the planet's  Bondi  radius  is  much smaller  than  the
softening length, the  stagnation point has an almost  fixed and large
value, so that it resides far  from the planet, whereas it lies within
the  Bondi  radius  when  the  latter is  larger  than  the  potential
softening  length. The  departure from  linearity therefore  occurs at
lower mass in the smaller softening length case.
Assuming  that the horseshoe  zone separatrix  does not  intersect any
shock (a  reasonable assumption for  small mass planets), one  can use
the invariance of  the Bernoulli constant in the  corotating frame, in
the steady state, to relate the perturbed quantities at the stagnation
point to the horseshoe zone width. The Bernoulli constant reads:
\begin{equation}
\label{eq:bernou}
J=\frac{u^2+r^2(\Omega-\Omega_p)^2}{2}+\Phi-r^2\Omega_p^2/2+\eta.
\end{equation}
This expression reduces,  at a stagnation point located  on the orbit,
to:
\begin{equation}
\label{eqn:jstag}
J_{\rm stag}=\Phi_*(a)+\eta_0+\tilde\Phi_S'-a^2\Omega_p^2/2,
\end{equation}
while it reads
\begin{equation}
\label{eqn:jsep}
J_{\rm sep}=(a+x_s)^2[\Omega(a+x_s)-\Omega_p]^2/2+\Phi_*(a+x_s)+\eta_0
-(a+x_s)^2\Omega_p^2/2
\end{equation}
on the separatrix, far from  the planet, where the effective potential
essentially   reduces  to   its  unperturbed   value  $\Phi_*+\eta_0$.
Equating  Eqs.~(\ref{eqn:jstag})  and~(\ref{eqn:jsep})  and  expanding
Eq.~(\ref{eqn:jsep}) to second order in $(x_s/a)$ yields:
\begin{equation} 
x_s=\frac1{\Omega_p}\sqrt{-\frac 83\tilde\Phi_S'}.
\end{equation} 
The horseshoe zone half width is therefore simply related to the value
of the Bernoulli constant at  the stagnation point.  We can understand
the boost  of the horseshoe region  width in the  transition region as
follows:
\begin{itemize}
\item  as long as  the flow  remains linear,  the stagnation  point is
  located  at a  fixed  position  far from  the  planet. It  therefore
  samples  a value  of the  perturbed Bernoulli  constant  that simply
  scales  with  $q$,  hence  the  horseshoe  zone  width  scales  with
  $q^{1/2}$.
\item  When $r_B\sim\epsilon$,  the  stagnation point  begins to  move
  towards  the planet  (see  Fig.~\ref{fig:diffbondi}), which  implies
  that $|\tilde\Phi_S'|/q$  is no longer a constant  but increases with
  $q$,  as  the  stagnation  point  goes  deeper  into  the  effective
  potential well  of the planet.  As a consequence the  horseshoe zone
  width increases faster than $q^{1/2}$ in this regime.
\end{itemize}

The  above  discussion is  valid  for a  2D  situation  with a  finite
potential   softening   length.    Under  these   circumstances,   the
dimensionless   parameter  that   controls  the   flow   linearity  is
$\epsilon/r_B$. In a three dimensional case with a point-like mass, we
can gain some insight on the condition for the flow linearity assuming
a horizontal, layered motion for each slice of disk material. Although
we  know that  this  is not  strictly  the case  \citep{dkh03}, it  is
nevertheless a useful approximation that relates the three dimensional
case to the  above discussion. In each slice,  the planet potential is
the one  of a  2D situation with  a potential softening  length $|z|$,
where $z$ is the slice altitude. Therefore, if over most of the disk's
vertical extent,  the flow  is linear  (that is, if  over most  of the
disk's vertical  extent, $|z|\gg r_B$, which amounts  to the condition
$H\gg r_B$) then  most of the torque acting on  the planet arises from
slices which contribute linearly to the torque, hence the total torque
nearly amounts to  its linearly estimated value, whereas  if the Bondi
radius  amounts  to a  significant  fraction  of  the disk's  vertical
extent, the layers with altitude $|z|<r_B$ have an excess of horseshoe
zone  width and contribute  significantly to  the total  torque value,
which   therefore  has   a   significant  offset   w.r.t  the   linear
estimate. The condition for the appearance  of the offset in a 3D case
is therefore $r_B\sim H$, which  also reads $q\sim h^3$, or using, the
notation of \citet{kpap96}, ${\cal M}\sim 1$.  This is consistent with
the dimensional  analysis of \citet{kpap96}  and with our  findings of
section~\ref{sec:depthick}.  We make the following comments:
\begin{itemize}
\item Although  the Bondi  sphere and the  Hill sphere  have different
  expression and scaling with the planet mass, they happen to coincide
  with the  disk thickness at roughly  the same planet  mass (within a
  factor  of $3$), so  that characterizing  the flow  non-linearity by
  comparing  the Hill  radius to  the disk  thickness also  amounts to
  comparing the Bondi radius to the disk thickness.
\item  Although we  probably do  not have  a sufficient  resolution to
  properly  characterize the flow  within the  Bondi radius  (when the
  softening length is  shorter than this radius), it  seems that there
  is no trapped  region of material librating about  the planet within
  this radius.  Indeed, in  Fig.~\ref{fig:flowtopo}B or C, we see that
  the  unique stagnation point,  within the  Bondi radius,  splits the
  disk material in its vicinity into four regions: the inner and outer
  disk,  and the  two ends  of the  horseshoe region.   This  may have
  important  consequences for  the numerical  simulations  of embedded
  planets in non self-gravitating disks:  in such disks, a common (and
  still debated) practice consists  in truncating the torque summation
  so as to reject  the contributions from the circumplanetary material
  \cite[e.g.][]{mp03}, which is considered  to form, together with the
  planet,  a  relevant  system  that  migrates as  a  whole,  and  the
  migration of  which is  accounted for by  the {\em  external} forces
  applied (hence the  truncation). In the case of  embedded small mass
  planets   however,  should   it   be  confirmed   that  no   trapped
  circumplanetary  material exists  in  the planet  vicinity, then  no
  torque truncation should be performed when evaluating the torque.
\item The  offset displays a remarkable amplitude  in 3D calculations,
  not even reproduced with  the relatively small softening length that
  we  adopted in  our  2D calculations  ($\epsilon=0.3H$). A  possible
  explanation for this is the  vertical motion of the disk material in
  the planet  vicinity described by \citet{dkh03}, which  results in a
  bent of the horseshoe streamlines  towards the planet.  As a result,
  the  stagnation point  associated to  the horseshoe  separatrix with
  altitude   $z$  far   away   from  the   planet   has  an   altitude
  $|z_s|<|z|$.  Therefore,  this stagnation  point  is  closer to  the
  planet   than  it   would  be   in  a   sliced   horizontal  motion
  approximation, hence the perturbed  Bernoulli constant at that point
  is larger  than given by  the horizontal motion  approximation, and
  the  associated horseshoe  separatrix  is wider,  yielding a  larger
  contribution to the coorbital corotation torque.

\end{itemize}
\section{Discussion}
\label{sec:discuss}
\subsection{Consequences for planetary migration}
\begin{figure}
\plottwo{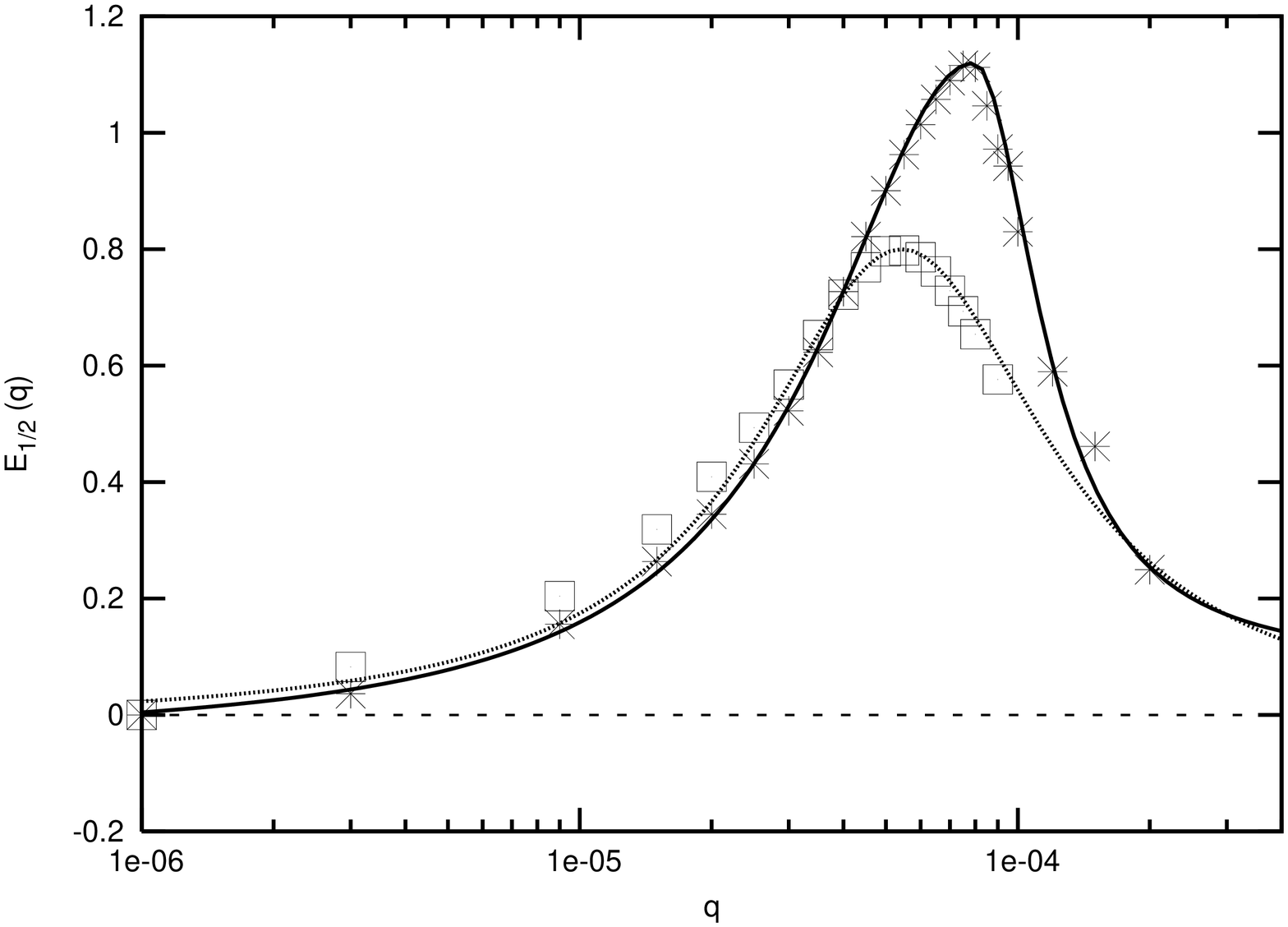}{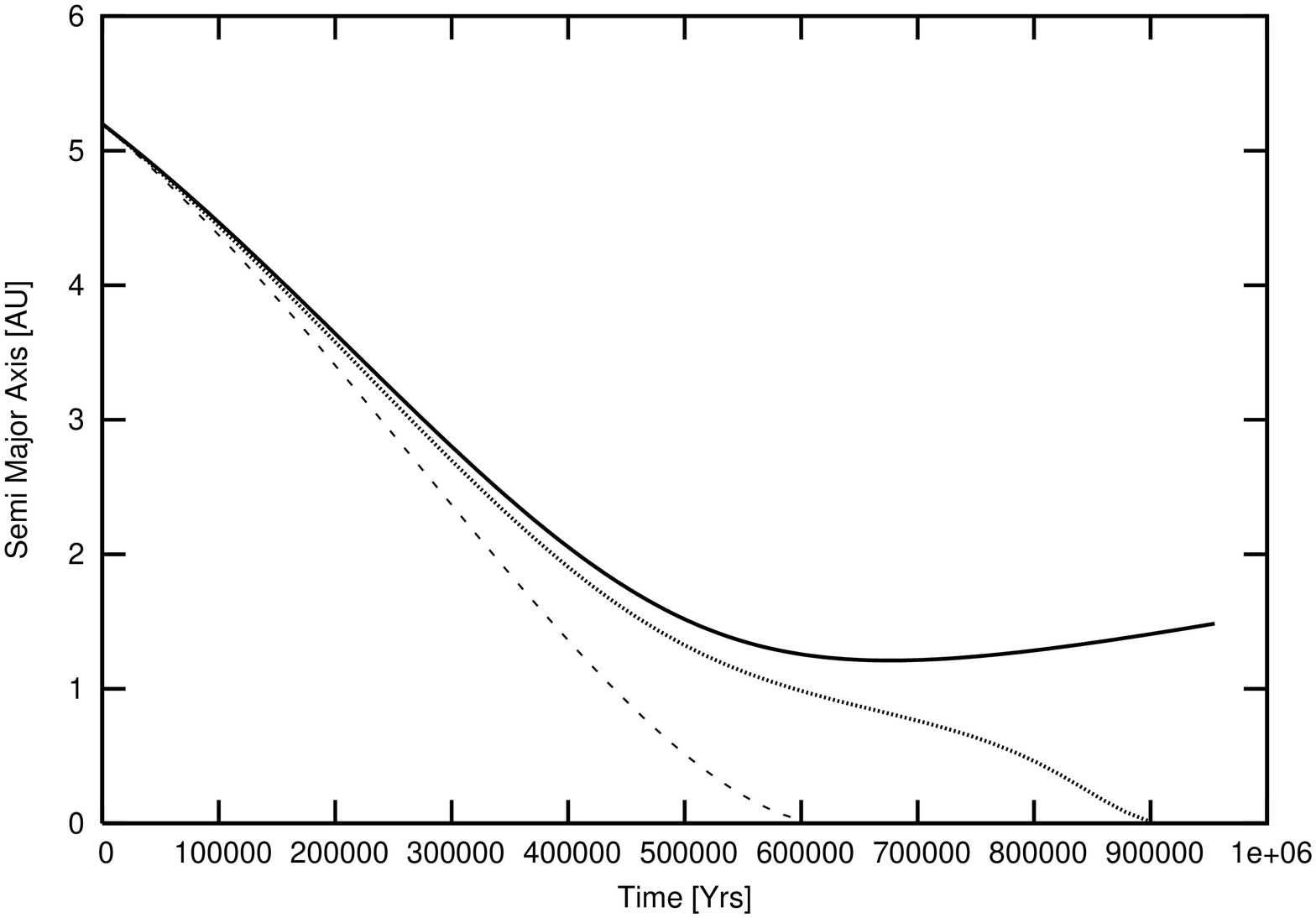}
\caption{\label{fig:evolution} Left: The  departure from linearity for
  3D  models with  $\alpha  =  1/2$ for  the  unsaturated (stars)  and
  saturated  (squares)  case.  The  solid  and dashed  lines  are  the
  corresponding  analytical   fit  formulae  used   for  evolving  the
  planet.  The   dashed  line  ($E_{1/2}=0$)  refers   to  the  linear
  case. Right: The  evolution of an embedded planet  in the disk using
  these  analytical formulae  (solid: unsaturated,  dotted: saturated,
  dashed: linear).}
\end{figure}
To  analyze the  effect of  the torque  offset from  linearity  on the
evolution  of  planets  in disks  we  have  performed  a set  of  test
simulations.   We start  from the  linear relation  for the  change in
semi-major axis of a planet as given by \citet{tanaka} for the 3D case
which can be written in the following form
\begin{equation}
 \dot{a}_{Lin}  = \, - \, 2 ( 1.364 + 0.541 \alpha) \, 
  \frac{\Sigma a^2}{M_*} \,  
   \, q \, a\, \Omega_p / h^2,
\end{equation}
which, in our system of units, can be recast as
%% WK
\begin{equation}
\label{eq:dota}
 \dot{a}_{Lin}  = \, - \, 2 ( 1.364 + 0.541 \alpha) \,
  \Sigma_0 \,
   \, q \, a^{3/2-\alpha} / h^2,
%%  \left[\frac{\mbox{AU}}{\mbox{yrs}}\right],
\end{equation}
where $\dot{a}_{Lin}$ is now given in units of AU/yrs, and in which we
used $M_*=1\,M_\odot$ and $r_0=1$~AU. In Eq.~\ref{eq:dota}, $\Sigma_0$
denotes the surface density at $r_0$ in units of $M_*/r_0^2$.
%%  End WK
To  model
deviations from linearity the above $\dot{a}_{Lin}$ is modified by our
numerically    found    offset    $E_\alpha(q)$    as    defined    by
Eq.~(\ref{eqn:ealpha}), while  the scaling  law for the  critical mass
$q_c \propto  h^3$ (cf. Sect.~\ref{sec:depthick}) is  included, in the
following manner:

\begin{equation}
\dot a= \dot a_{Lin}\left[1-E_\alpha\left(\frac{qh_0^3}{h^3}\right)\right],
\end{equation}
where $h_0=0.05$  is the disk aspect  ratio for which  we have sampled
the dimensionless offset $E_\alpha(q)$ by 3D calculations.

To make the simulations numerically simpler the hydrodynamically found
data points are approximated by  analytical functions, where we find a
combination  of two  Lorentzians matched  at $q=q_c$  very  useful. In
addition we use,  for demonstration only, a linear  growth law for the
planetary mass $q  = q_0 \, t/t_{grow}$.  To  integrate the equation a
standard 4$^{\rm th}$ order Runge-Kutta scheme is used.

As  an illustrative  example  we have  performed  simulations for  the
intermediate case $\alpha =  1/2$, and in Fig.~\ref{fig:evolution} our
results are displayed. The left panel shows the offset for the 3D case
for  the unsaturated and  partially saturated  torques ($\nu=10^{-5}$)
with $h=0.05$,  where the symbols  refer to the  hydrodynamical models
described above  and the lines  refer to the analytical  fit formulae.
In the right panel we display our results on the migration of a planet
in the presence of an offset from linearity, using $q_0 = 10^{-5}$ and
$t_{grow} =  10^5$yrs. For the flaring  of the disk we  use $h \propto
r^{0.28}$  with  $h=0.07$  at  $r=5.2$AU  and a  value  of  $\Sigma  =
300$g/cm$^2$
%% WK
at $r_0=1$AU, translating to $\Sigma_0 = 3.4 \times 10^{-5}$.
%% End WK

The dashed line refers to the standard linear case, the dotted line to
the partially  saturated case, and  the solid line to  the unsaturated
case.  Clearly the offset yields an extended migration time scale.  In
the partially saturated case,  where $E_\alpha$ remains always smaller
than unity, the total migration  time (to reach $r=0$) is increased by
roughly 50\%. In the unsaturated case, where $E_\alpha$ is larger than
unity at  the critical $q_c$, we  find indeed a {\it  reversal} of the
migration.   This is  possible if  during the  migration process  of a
planet the local $h(r)$ is such  that the actual mass of the planet is
above the minimal mass for  migration reversal [i.e.  the mass $q_{\rm
  min}$ for which $E_\alpha(q_{\rm min})=1$].

\begin{figure}
\plotone{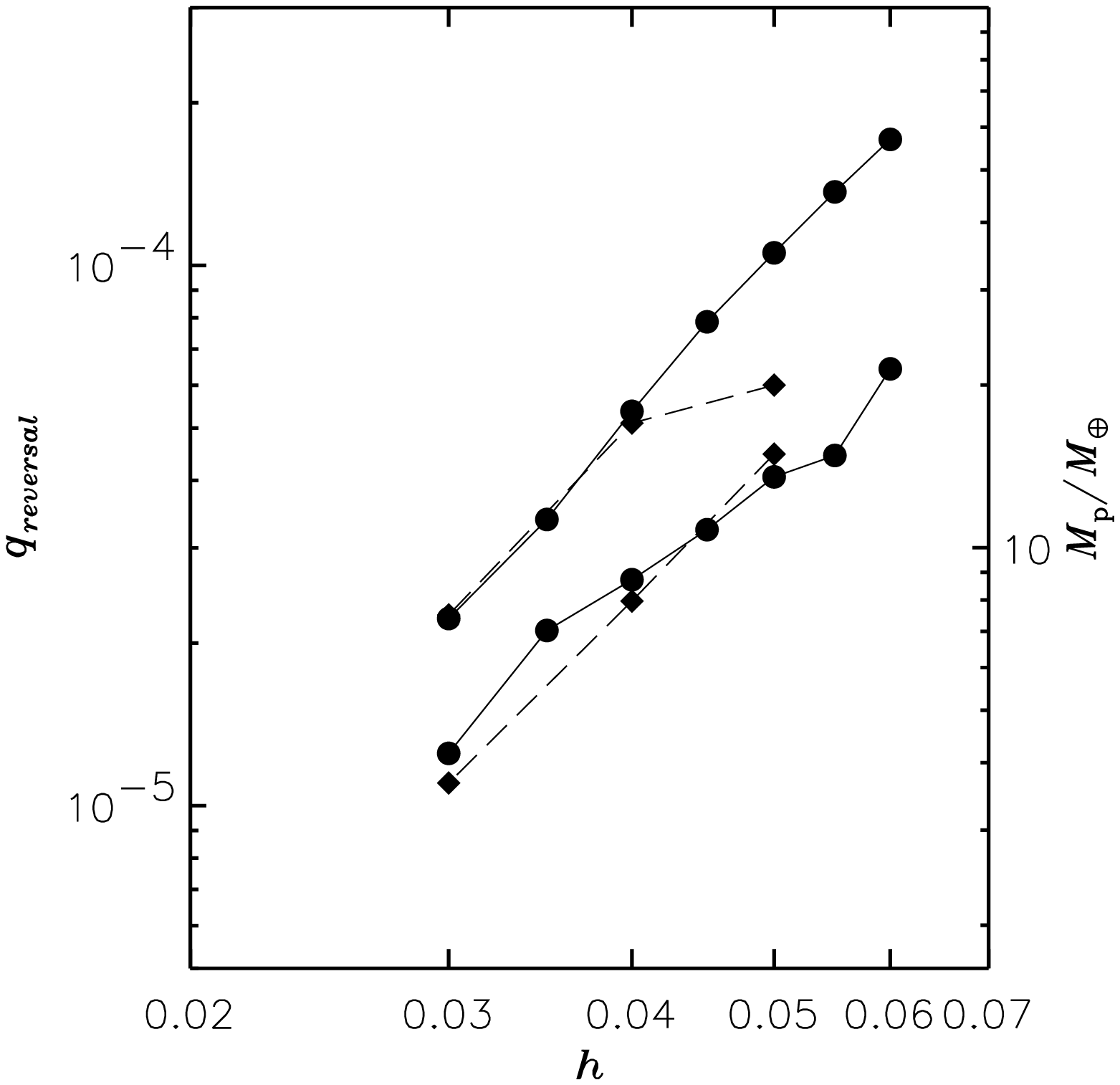}
\caption{\label{fig:reversal}  Domain of  migration  reversal, in  the
  $(h,q)$-plane, in the unsaturated  case (solid curves) and partially
  saturated  case  (dashed curve).   For  each  line  style (solid  or
  dashed), the  lower curve represents the minimal  mass for migration
  reversal  while the  upper  curve represents  the  maximal mass  for
  migration  reversal.  At  low  $h$ (hence  low  $q$), the  partially
  saturated  and  unsaturated   results  almost  coincide,  since  the
  corotation  is  very  weakly  saturated  (we work  with  a  constant
  kinematic viscosity),  while the reversal  domain is more  narrow at
  large $h$, owing to the increasing corotation torque saturation. The
  right axis labeling assumes a solar mass central object.}
\end{figure}

We have also thoroughly  investigated the migration reversal domain in
the flat  surface density case ($\alpha=0$), for  the unsaturated case
(short   runs)  and   partially   saturated  case   (long  runs   with
$\nu=10^{-5}$).      The       results      are      displayed      in
Fig.~\ref{fig:reversal}. In this figure  one can see that the reversal
domain,   for   $h=0.03-0.05$,   typically   corresponds   to   masses
representative of sub-critical solid cores of giant planets.
\subsection{Corotation torque saturation issues}
As we  already mentioned in section~\ref{sec:viscdep},  in the absence
of  any process  that  allows angular  momentum  exchange between  the
horseshoe region  and the rest  of the disk, the  coorbital corotation
torque  saturates after a  few libration  timescales \citep{bk01,m02}.
Such  exchange cannot  be provided  by pressure  waves excited  by the
planet, as these  wave corotate with the planet  and are evanescent in
the coorbital  region. The viscous  stress at the separatrices  of the
horseshoe region  gives rise  to a net  flux of angular  momentum from
this region to the inner or  outer disk.  In principle, some amount of
disk  viscosity should  therefore be  able to  prevent  the corotation
torque saturation.   An estimate of the minimum  viscosity required to
prevent  the  torque saturation  can  be  determined  as follows:  the
saturation results from the libration,  which tends to flatten out the
vortensity  profile across  the horseshoe  region (in  an  inviscid 2D
flow, the vortensity  is conserved along a fluid  element path), while
viscous  diffusion  tends  to   restore  the  large  scale  vortensity
gradient, if  any. It  succeeds in doing  so if the  viscous timescale
across the  horseshoe region is  shorter than the  libration timescale
\citep{wlpi92,m01,m02}. This yields:
\begin{equation}
\label{eqn:viscsat}
\nu_m=0.035\left(\frac qh\right)^{3/2}a^2\Omega_p,
\end{equation}
where $\nu_m$ is  the minimal viscosity to avoid  the coorbital torque
saturation \citep{mmcf06}. As  can be seen in Eq.~(\ref{eqn:viscsat}),
it is easier to desaturate the corotation torque of lower mass planets
(the minimal viscosity  required to do so is  smaller). The reason for
this  is twofold:  as the  planet mass  decreases, the  horseshoe zone
width decreases, therefore (i)  the libration time increases, (ii) the
viscous timescale  across the  horseshoe region decreases.   Recast in
terms of an $\alpha$-parameter\footnote{In this section only, $\alpha$
  denotes in  a standard manner  the effective kinematic  viscosity in
  units of $H^2\Omega$, as introduced by \citet{ss73}, rather than the
  surface    density   slope    index,   as    previously   defined.},
Eq.~(\ref{eqn:viscsat}) reads:
\begin{equation}
\label{eqn:alphamin}
\alpha_m=0.035q^{3/2}h^{-7/2},
\end{equation}
We can  use the fact  that the  mass ratio $q$  at the maximum  of the
offset is a linear function of $h^3$, that reads:
\begin{equation}
\label{eqn:qhrel}
q\approx 0.56h^3,
\end{equation}
as   can  be   easily  found   from   Fig.~(\ref{fig:3ddeph}).  Using
Eq.~(\ref{eqn:qhrel}) to substitute either $h$ or $q$ in
Eq.~(\ref{eqn:alphamin}), we obtain either:
\begin{equation}
 \alpha_m\approx 0.018q^{1/3},
\end{equation}
or
\begin{equation}
 \alpha_m\approx 0.015h.
\end{equation}
These equivalent  expressions give  the minimal viscosity  required to
prevent the saturation of the  corotation torque for a planet mass for
which  the offset  is  maximal,  i.e.  for  which  migration could  be
significantly slowed down or  reversed, provided the corotation torque
amounts to a sizable fraction of its unsaturated value. In a disk with
$h=0.04$, this  yields: $\alpha_m=6\cdot 10^{-4}$, which  falls in the
range of  the $\alpha$  values inferred from  observations of  T Tauri
stars, for which $\alpha=10^{-4}-10^{-2}$.

The molecular viscosity of the  gas is however orders of magnitude too
low to account for such  values of $\alpha$.  It is generally admitted
that  a large  fraction of  a protoplanetary  disk is  subject  to the
magnetorotational instability or  MRI \citep{balbh91a}, the non-linear
outcome of  which is a turbulent  state which endows the  disk with an
effective  kinematic  viscosity  of  the  order of  magnitude  of  the
viscosity needed to account for  the mass accretion rate inferred from
observations of  T Tauri  disks.  In such  disks, however,  the torque
exerted by the gas on  an embedded protoplanet displays large temporal
fluctuations that tend to yield a random walk of the planet semi-major
axis,    rather    than    a    steady    drift    of    the    latter
\citep{np04,nelson05}. \citet{nelson05} has shown that even for planet
masses of  the order of  $10-30$~$M_\oplus$ (in a disk  with $h=0.07$,
with  no  vertical stratification),  the  random  fluctuations of  the
semi-major axis overcome the effects of type~I migration on timescales
of the order of  $O(10^2)$~orbits, while \citet{jgm06} argue that such
diffusive migration  systematically lowers the  planet lifetimes, even
if it allows a small fraction of protoplanets to ``survive'' migration
over the disk lifetime. In  MHD turbulent disks, the stochastic nature
of the  turbulent viscosity,  although largely sufficient  to maintain
the  corotation torque  unsaturated, would  certainly hide  the effect
that we describe in this work, at least over $O(10^2)$ orbits.  Should
the random fluctuations average out  over longer timescales, so that a
systematic drift  could be reliably measured, the  effect of migration
slow down of sub-critical solid cores should become noticeable%
\footnote{Provided that the total torque,  in a turbulent disk, can be
  considered as  the sum of  the fluctuations arising  from turbulence
  and of the laminar torque, which remains to date an open question.}.

There are  other situations, yet numerically unexplored,  in which the
disk's turbulent  state could  prevent the corotation  torque saturation
and  yet  be  sufficiently  mild  that  the  planet  would  undergo  a
systematic rather than stochastic migration. This could be the case of
the so-called dead zone, a region of the disk where the gas ionization
fraction is too  low to allow the coupling of the  gas to the magnetic
field and where the MRI does not occur.  The disk upper layers above a
dead zone  are sufficiently ionized by external  irradiation of cosmic
rays or high-energy photons to be  subject to the MRI and therefore to
be  turbulent   \citep{ga96}.   This  turbulence   generates  velocity
fluctuations  at the  disk midplane,  within the  dead zone,  which is
therefore not  completely ``dead'' and  has an $\alpha$  value several
times     smaller     than    that     of     the    active     layers
\citep{fs03,RR03,fp06}. It  is likely that  within the dead  zone, the
torque convergence  is reached, over  a given timescale, at  a smaller
planet  mass  than in  an  MHD  turbulent  disk, which  suggests  that
sub-critical   solid   cores  could   undergo   a  steady   migration,
significantly slowed down, or reversed, within the dead zone.

It is also possible that weaker forms of turbulence may exist that are
still able  to prevent the  corotation torque saturation, such  as the
hydrodynamics   turbulence   triggered   by  the   global   baroclinic
instability \citep{kb03}.  However, the turbulence  resulting from the
Kelvin  Helmholtz instability due  to the  gas vertical  shear arising
from  the dust  sedimentation \citep{jhk06}  seems to  be too  weak to
desaturate the  corotation torque for  planet masses  larger than
$\sim 1\;M_\oplus$,  as it  yields an $\alpha$-value  of the  order of
$10^{-6}$.

We close this  section with the following comment:  all what is needed
to  avoid  the corotation  torque  saturation  is  to bring  ``fresh''
vortensity from  the inner  or  outer disk  to the  horseshoe
region in less than a libration timescale. The standard approach based
upon the comparison of the libration and viscous timescales across the
horseshoe region is certainly correct when the largest turbulent scale
is  smaller  than the  horseshoe  zone  width,  so that  the  
vortensity enters the horseshoe region in a diffusive manner, but it is
unlikely to be  adequate when the turbulence scale  is larger than the
horseshoe region width. In this case, which occurs among others in the
case of  the MHD turbulence, one  rather has to  compare the libration
timescale to  the advection timescale  across the horseshoe  region at
the average turbulent speed.  This plays in favor of desaturation, and
seems  to imply that  preventing the  corotation torque  saturation is
much  easier   than  suggested  by   the  libration/viscous  diffusion
timescales comparison.

\section{Conclusion}
\label{sec:conclusion} 
By means of two and three dimensional calculations we have found the
following:
\begin{enumerate}
\item  There  is  a  boost  of the  coorbital  corotation  torque  for
  sub-critical solid cores ($M\la15\;M_\oplus$) in thin ($H/r\la0.06$)
  protoplanetary  disks.   In   disks  with  shallow  surface  density
  profiles, i.e.  $\Sigma(r)\propto  r^{-\alpha}$ with $\alpha < 3/2$,
  this yields  a {\em positive}  excess of the corotation  torque that
  leads to a slowing down or reversal of the migration.
\item This  boost appears  to be the  first manifestation of  the flow
  non-linearity (prior  to gap opening, which occurs  at larger planet
  mass).
\item The horseshoe  region has a width that  scales as $M_p^{1/2}$ at
  low planet mass (linear regime), whereas it scales as $M_p^{1/3}$ at
  large planet  mass.  At the  transition between the two  regimes the
  horseshoe region is wider  than linearly predicted, which yields the
  aforementioned boost of the corotation torque.
\item Since this is a  non-linear effect, its occurrence is controlled
  by  the dimensionless parameter  ${\cal M}=R_H/H$,  or $r_B/H=3{\cal
    M}^3$. For a disk of given aspect ratio $h$, the corotation torque
  enhancement is maximal for a planet mass $M_p$ given by
\begin{equation}
M_p\approx5 \left(\frac{M_*}{M_\odot}\right)\left(\frac{h}{0.03}\right)^3\;M_\oplus,
\end{equation}
which represents a mass typical for solid cores of giant protoplanets,
those for which  the (type I) migration timescale  problem is the most
acute.
\item The torque reversal, if any, occurs therefore at lower masses in
  thinner disks (lower aspect ratio).  As a consequence, the migration
  of  a planet  of given  mass would  stop, in  a flaring  disk,  at a
  distance  from  the  central  object  that  depends  on  the  planet
  mass. Conversely,  if an accreting  protoplanet, in a  flaring disk,
  reaches a  point where  the tidal torque  cancels out, it  starts to
  recede from the central object at a rate dictated by its mass growth
  rate.
\item  This effect  has been  unnoticed  thus far  in 2D  calculations
  probably owing  to the large  softening length adopted or  to strong
  torques   arising  from   within   the  Roche   lobe  of   accreting
  planets. Poor mass sampling may have also played a role.
\item Small  mass planets do not  have a Roche lobe  (i.e.  a prograde
  circumplanetary  disk   extending  over  a  fraction   of  the  Hill
  radius). They have a Bondi  sphere, that is smaller than their Roche
  lobe.  There  is presently an  issue about the torque  evaluation in
  calculations   with   non    self-gravitating   disks.    In   these
  calculations, it is still debated whether one must include the Roche
  lobe content  \citep{dbl05} or  not \citep{mp03} in  the sum  of the
  elementary  contributions  to  the  torque  of  the  disk  material.
  Regardless of  the correct answer to this  question, numericists who
  truncate the torque summation in the planet vicinity should be aware
  that the  sum should only exclude  at most the  (small) Bondi sphere
  rather than the Roche  lobe when simulating deeply embedded ($r_B\ll
  H$) protoplanets.
\item In 2D calculations,  the dimensionless parameter that determines
  the  flow linearity in  the planet  vicinity is  $r_B/\epsilon$.  If
  $\epsilon\propto H$  (a prescription that  we chose for the  2D runs
  presented in this work) or $\epsilon\propto R_H$, this dimensionless
  parameters   scales  as  a   function  of   $q/h^3$  and   the  flow
  non-linearities in such 2D  calculations also appear for mass ratios
  $q\propto h^3$.
\end{enumerate}
We suggest that  the findings listed above could  motivate future work
on the following points:
\begin{enumerate}
\item The flow  transition exhibited in this work  could be studied in
  the simplified  framework of the shearing  sheet approximation. Then
  the asymmetry between the left and right stagnation points (which we
  believe to be a feature of minor importance, despite its robustness)
  would disappear,  and they  would lie on  the same  separatrix. This
  study  could be  undertaken using  the method  of  \citet{kpap96}. A
  quantitative study  of the flow  transitions (planet mass  for which
  the left and  right stagnation points coalesce, and  planet mass for
  which a Roche lobe appears) would provide a very valuable insight on
  the dynamics of the flow in the planet vicinity.
\item Although we have seen that in the low mass case (deeply embedded
  core, or  $r_B\ll H$) the  flow non-linearities are confined  to the
  Bondi sphere, we  do not have undertaken a study  of the flow within
  this  sphere. Characterizing this  flow, possibly  by means  of very
  high  (nested  grid)  numerical   simulations,  would  be  of  great
  interest.
\item  The  role  of  accretion  has been  neglected  in  the  present
  analysis, while  the mass  range for which  the offset  is observed,
  depending  on the disk  thickness, may  involve accreting  cores. It
  seems  that  accretion   enhances  the  offset \citep{dkh03},  but  a
  quantitative analysis of its impact remains to be done.
\item We have emphasized the role played by dissipation, which must be
  present to prevent the  corotation torque saturation. As the present
  study deals with small mass planets, it should be relatively easy to
  prevent  this   saturation.   So   far  the  only   self  consistent
  calculations of a  turbulent disk with embedded planets  deal with a
  fully turbulent disk subject to  the MRI. A study characterizing the
  ability of other forms of  turbulence (such as the global baroclinic
  instability,  or the  residual  turbulence  of the  dead  zone in  a
  layered accretion disk) to desaturate  the corotation torque of small
  mass planets would be very valuable.
\end{enumerate}

\acknowledgments The computations with  NIRVANA reported in this paper
were performed using the UK Astrophysical Fluids Facility (UKAFF). The
computations  with FARGO  were performed  at the  Centre de  Calcul de
l'IN2P3. GD  acknowledges support from the Leverhulme  Trust through a
UKAFF Fellowship, from  the NASA Postdoctoral Program, and in part from
NASA's Outer Planets Research Program through grant 811073.02.01.01.20.
The authors are indebted  to Hidekazu Tanaka  for bringing to their  
attention the role played by the Bondi radius.

\end{document}